\documentclass[aps,pra,twocolumn,showpacs,longbibliography]{revtex4-2}
\usepackage{graphicx} %Include figure files,superscriptaddress
\usepackage{amsmath}
\usepackage{graphicx,epstopdf}
\usepackage{gensymb}
\newcommand{\be}{\begin{equation}}
	\newcommand{\ee}{\end{equation}}
\newcommand{\bea}{\begin{eqnarray}}
	\newcommand{\eea}{\end{eqnarray}}
\newcommand{\bse}{\begin{subequations}}
	\newcommand{\ese}{\end{subequations}}

\usepackage{color}
\usepackage[colorlinks,bookmarks=false,citecolor=darkblue,linkcolor=red,urlcolor=blue]{hyperref}
\usepackage{multirow}

\definecolor{darkred}{rgb}{0.7,0.0,0.0}

\definecolor{darkblue}{rgb}{0,0.02,0.45}

\definecolor{darkgreen}{rgb}{0.02,0.45,0.0}

\definecolor{violet}{rgb}{0.8,0.2,0.6}

\begin{document}

\title{Magnetic and crystal electric field studies of two Yb$^{3+}$-based triangular lattice antiferromagnets}
\author{S. Guchhait}
\author{R. Kolay}
\author{A. Magar}
\author{R. Nath}
\email{rnath@iisertvm.ac.in}
\affiliation{School of Physics, Indian Institute of Science Education and Research Thiruvananthapuram-695551, India}
%\date{\today}

\begin{abstract}
We present the low-temperature magnetic properties of two Yb$^{3+}$-based triangular lattice compounds NaSrYb(BO$_3$)$_2$ and K$_3$YbSi$_2$O$_7$ via thermodynamic measurements followed by crystal electric field (CEF) calculations. Magnetization and specific heat data as well as the CEF energy levels confirm that the ground state is characterized by the low-lying Kramers' doublet of Yb$^{3+}$ with effective spin-1/2 ($J_{\rm eff} = 1/2$). A small Curie-Weiss temperature and scaling of magnetic isotherms corroborate very weak magnetic correlations among $J_{\rm eff} = 1/2$ spins.
The crystal field parameters are calculated using the point charge model and the CEF Hamiltonian is determined for both the compounds. The simulation using the eigenvalues of the CEF Hamiltonian reproduces the experimental susceptibility, magnetic isotherm, and magnetic specific heat data very well. The large separation between the ground state and first excited state doublets implies that the ground state is a Kramers' doublet with $J_{\rm eff} = 1/2$ at low temperatures, endorsing the experimental findings.
\end{abstract}

\maketitle
\section{\textbf{Introduction}}
Quasi-two-dimensional (2D) geometrically frustrated magnets with low spin value (e.g. $S = 1/2$) have strong quantum fluctuations that melt the conventional magnetic
long-range order (LRO), leading to several disordered ground states, such as, quantum spin liquid (QSL)~\cite{Savary016502,*Balents199}. Spin-$1/2$ triangular lattice antiferromagnet (TLA) is a simplest example of the geometrically frustrated lattice in 2D, was first proposed by Anderson to host resonating valence bond state, a prototype of QSL~\cite{Anderson153}.
A detailed theoretical study of an isotropic Heisenberg TLA for arbitrary $S$-value has suggested to have a non-collinear 120$^{\degree}$ ordered state in zero-field~\cite{Singh1766,Capriotti3899}. Subsequently, a series of theoretical and numerical studies on $S = 1/2$ TLA with anisotropic nearest-neighbor (NN) and next-nearest-neigbbour (NNN) Heisenberg interactions have revealed numerous interesting quantum phases at the critical ratios of the exchange couplings~\cite{Zhu207203,Drescher220401,Zhu041105,Hu140403}. Moreover, atomic disorder and vacancies are often proven to be pertinent in regards to stabilizing QSL~\cite{Kimchi031028,Li107202,Kundu117206,Zhu157201}. Therefore, relentless efforts are being made in order to design suitable geometrically frustrated magnets which may allow to probe the intriguing ground states.

Recently, rare-earth ($4f$)-based TLAs, especially with Yb$^{3+}$ ion, provide a new platform to investigate exotic quantum phases of matter. In contrast to transition metal oxides, the interplay between strong spin-orbit coupling (SOC) and weak crystal electric field (CEF) in $4f$ systems leads to a Kramers' doublet with an effective spin $J_{\rm eff} = 1/2$ ground state at low temperatures~\cite{Somesh104422,Lal014429}. For instance, well-studied compounds YbMgGaO$_4$ and chalcogenides NaYb$C_2$ ($C$ = O, S, Se) with $J_{\rm eff} = 1/2$ ground state are reported to feature QSL and field-induced ordered states~\cite{Li097201,Dai021044,Li15814,Sarkar241116,Scheie014425}. Furthermore, rare-earth-based frustrated magnets with reduced exchange couplings are potential materials for achieving sub-Kelvin temperatures via
adiabatic demagnetization refrigeration (ADR) technique~\cite{Treu013001}. Recently, a series of rare-earth-based TLAs with general formula $AA^{\prime}R$(BO$_3)_2$ (where, $A$ = K, Rb, Na; $A^{\prime}$ = Ba, Sr; and $R$ = rare-earth ions) have demonstrated several exotic ground states and proven to be good ADR materials to reach mili-Kelvin temperatures. KBaYb(BO$_3)_2$ and KBaGd(BO$_3)_2$ are two well studied systems in this family~\cite{Tokiwa42,Jesche104402}. KBaYb(BO$_3)_2$ exhibits no magnetic LRO due to the geometrical frustration and site disorder of K$^{+}$ and Ba$^{2+}$ ions~\cite{Tokiwa42}. On the other hand, KBaGd(BO$_3)_2$ undergoes a magnetic LRO at around $T_{\rm N} \simeq 263$~mK in zero magnetic field. Using ADR technique, one can achieve temperatures as low as $\sim 122$~mK and 40~mK for KBaGd(BO$_3)_2$ and KBaYb(BO$_3)_2$, respectively~\cite{Jesche104402}. Interestingly, KBaGd(BO$_3)_2$ shows a dipolar spin-liquid phase between the antiferromagnetic order state and the paramagnetic region due to the interplay of dominant dipolar coupling and weak Heisenberg interaction~\cite{Xiang2023}.
%In QSL state robust quantum fluctuations prevent the conventional magnetic order state down to absolute temperature despite strong exchange interaction between electron spins. The spins of QSL state are highly entangled and support exotic fractional excitations that are necessary for quantum computing~\cite{Broholm0668}.

\begin{figure}
\includegraphics[width=\linewidth]{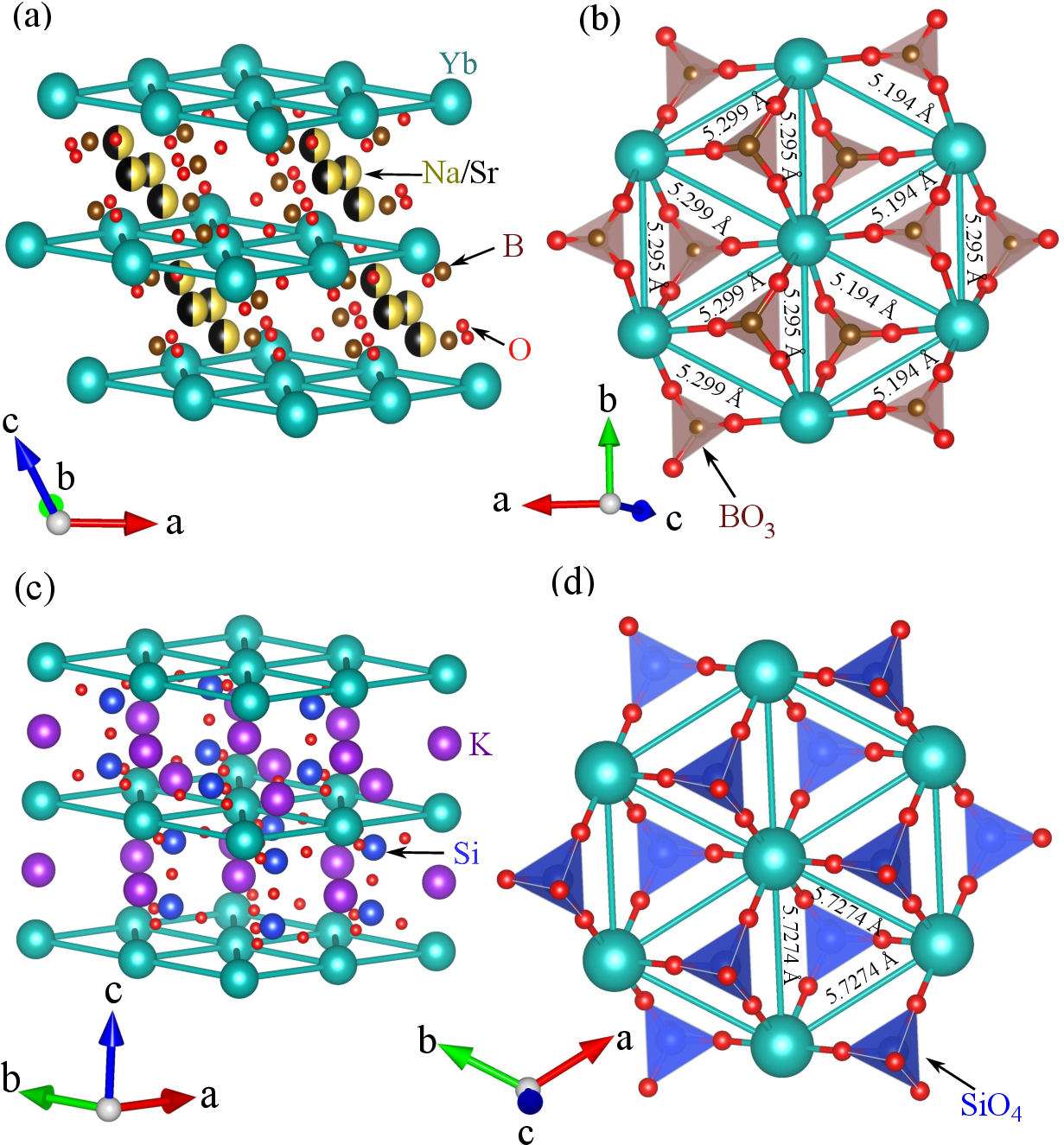} 
\caption{(a) Crystal structure of NaSrYb(BO$_3$)$_2$ showing triangular layers of Yb$^{3+}$ ions separated by Na and Sr (site mixing) atoms. (b) A section of the triangular layer showing Yb$^{3+}$ ions linked via BO$_3$ units on an anisotropic triangular lattice. (c) Crystal structure of K$_3$YbSi$_2$O$_7$. (d) A section of the isotropic triangular layer formed by Yb$^{3+}$ ions.}
\label{Fig1}
\end{figure}
%Several rare-earth-based triangular lattices compounds such as PrZnAl$_{11}$O$_{19}$~\cite{Bu134428}, YbMgGaO$_4$~\cite{Li097201}, and YbBO$_3$~\cite{Somesh064421}, kagome lattice compound as Pr$_3$GaSiO$_4$~\cite{Zorko057202, Lumata224416} and Nd$_3$GaSiO$_4$~\cite{Simonet237204}, which are reported to be useful for studying novel magnetic phases. 

Herein, we present the ground properties of two frustrated TLA compounds NaSrYb(BO$_3$)$_2$ and K$_3$YbSi$_2$O$_7$. NaSrYb(BO$_3$)$_2$ belongs to a family $AA^{'}R$(BO$_3$)$_2$ and crystallizes in a monoclinic structure with space group $P2_1/m$~\cite{Kuznetsov7497}. In the crystal structure, distorted YbO$_6$ octahedra are corner-shared with BO$_3$ triangles and put up an anisotropic 2D triangular lattice [see Fig.~\ref{Fig1}(b)] in the $ab$-plane. The triangular layers are well separated by the disordered Na$^{+}$ and Sr$^{2+}$ atoms [site (Na/Sr) mixing], as illustrated in Fig.~\ref{Fig1}(a). On the other hand, K$_3$YbSi$_2$O$_7$ crystallizes in a hexagonal structure with space group $P6_3/mmc$~\cite{Dabic584}. In K$_3$YbSi$_2$O$_7$, regular YbO$_6$ octahedra are corner shared through SiO$_4$ tetrahedra, forming an isotropic triangular layer in the crystallographic $ab$-plane [see Fig.~\ref{Fig1}(d)]. These triangular planes are connected via corner-sharing of two SiO$_4$ units through the apical oxygen along the crystallographic $c$-axis and the K$^{+}$ ions sit in-between two adjacent layers, as shown in Fig.~\ref{Fig1}(c). No conventional magnetic LRO is detected down to 0.4~K in both compounds. The crystal field calculations using the point charge model give a tentative estimation of the CEF energy levels of all expected doublets for both the compounds. Finally, we simulated the magnetic susceptibility, magnetic isotherms, and specific heat using the CEF parameters and made a comparison with the experimental data.

%We compare the end ADR temperature and volumetric entropy density of these compounds with different ADR magnets. 

%The Yb–Yb distances in the triangular layer varies in the range of 5.19~\AA to 5.30~\AA. The Yb–Yb distances of two adjacent layers are lies in the range of 6.40~\AA to 6.43~\AA and provides additional frustration because adjacent triangular layers are shifted relative to each other, as shown in Fig.~\ref{Fig1}(a).

\section{Methods}
\begin{figure}[h]
\includegraphics[width=\linewidth]{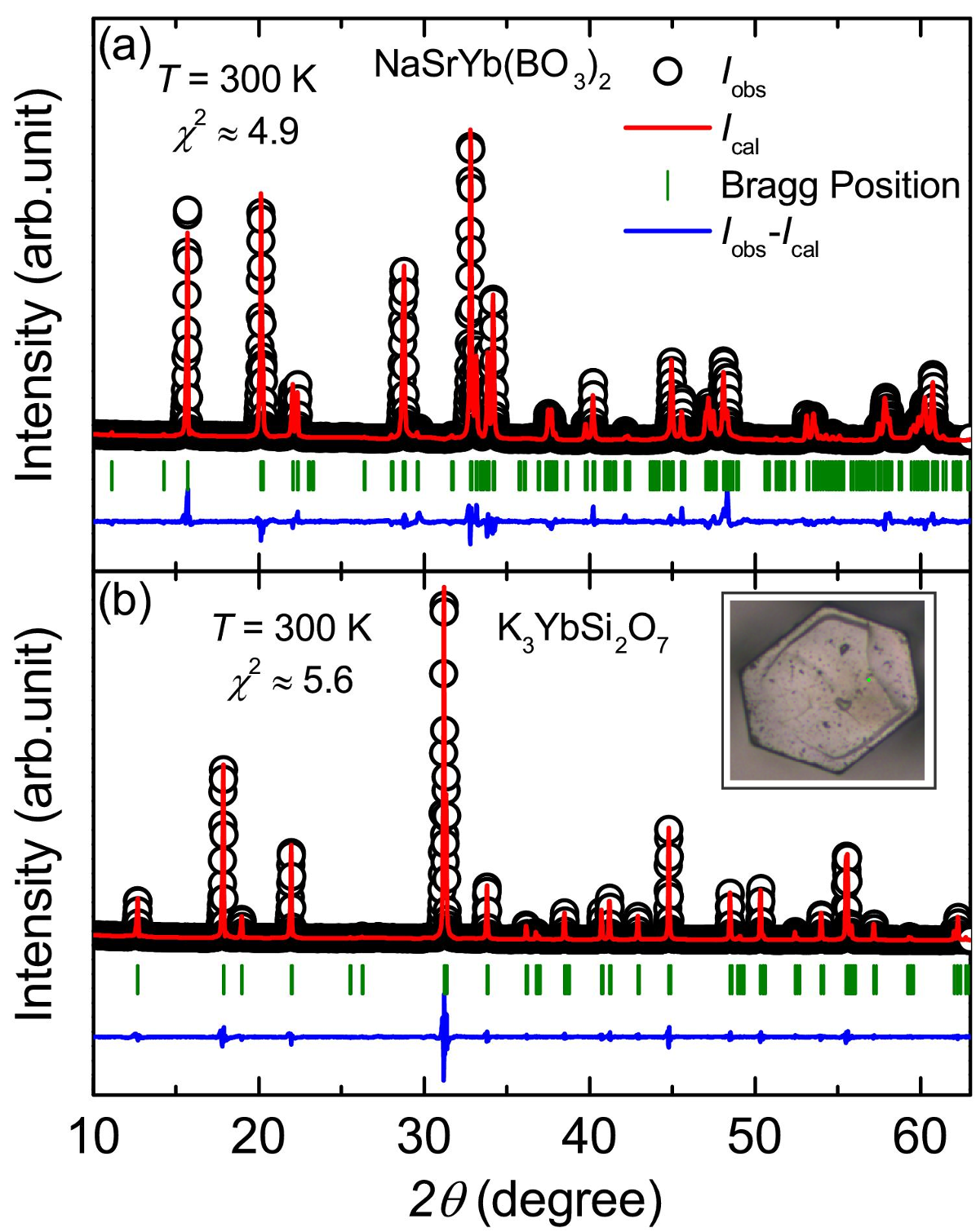}
\caption{Powder XRD patterns of (a) NaSrYb(BO$_3$)$_2$ and (b) K$_3$YbSi$_2$O$_7$ at room temperature. Red solid line represents the Rietveld fit. The black circles denote observed intensity. Green small vertical bars at the bottom show the expected Bragg peak positions and the lower blue solid line corresponds to the difference between observed and calculated intensities. $\chi^2$ represents the goodness-of-fit of the Rietveld refinement. The inset of (b) shows the image of a single crystal of K$_3$YbSi$_2$O$_7$.}
\label{Fig2}
\end{figure}

Polycrystalline sample of NaSrYb(BO$_3$)$_2$ was synthesized by the conventional solid-state reaction method in a platinum crucible. For its synthesis, stoichiometric mixture of Yb$_2$O$_3$ (Aldrich, 99.9\%), Na$_2$CO$_3$ (Aldrich, 99.9\%), SrCO$_3$ (Aldrich, 99.8\%), and H$_3$BO$_3$ (Aldrich, 99\%) was preheated at 650$^{\degree}$C for 5~hrs to decompose the starting materials. In the next step, the mixture was ground thoroughly, pressed into pellets, and annealed at 850$^{\degree}$C for 12~hrs. Phase purity of NaSrYb(BO$_3$)$_2$ was confirmed by powder x-ray diffraction (XRD) measurement at room temperature using a PANalytical x-ray diffractometer with Cu~$K_{\alpha}$ radiation ($\lambda_{\rm avg} \simeq 1.5418$~\AA). Rietveld refinement of the powder XRD pattern was performed using the FULLPROF software package~\cite{Carvajal55}, taking the initial structural parameters of the iso-structural compound NaSrY(BO$_3$)$_2$~\cite{Kuznetsov7497}. All the peaks were indexed properly using the monoclinic structure with space group $P2_1/m$ (No.~11) [see Fig.~\ref{Fig2}(a)]. The obtained lattice parameters and unit cell volume ($V_{\rm cell}$) are $a = 9.0605(4)$~\AA, $b = 5.2958(2)$~\AA, $c=6.4305(3)$~\AA, $\beta = 118.51(3)$, and $V_{\rm cell} \simeq 271.12$~\AA$^{3}$. The refined atomic coordinates of different atoms for NaSrYb(BO$_3$)$_2$ are listed in Table~\ref{Refined parameters}
\begin{table}[ptb]
	\caption{The Wyckoff positions and refined atomic coordinates for each atom of NaSrYb(BO$_3$)$_2$ at room temperature, obtained from the powder XRD.}
	\label{Refined parameters}
	\begin{ruledtabular}
		\begin{tabular}{ccccccc}	
			\multicolumn{1}{p{1cm}}{\centering Atomic \\ sites} & \multicolumn{1}{p{1.3
				cm}}{\centering Wyckoff\\ positions} &\multicolumn{1}{p{2
				cm}}{\centering $x$} &\multicolumn{1}{p{1.3
				cm}}{\centering $y$} &\multicolumn{1}{p{1.3
				cm}}{\centering $z$}  &  Occ.
			\\\hline
			Yb(1) & 2c &0.255(1) & 0.25& 0.499(2) & 0.5 \\
			Na(1) & 6g & 0.450(2) & 0.250 & 0.143(3) & 0.241 \\
			Sr(1) & 6g & 0.450(2) & 0.250 & 0.143(3) & 0.259 \\
			Na(2) & 6g & 0.0513(2) & 0.250 & -0.164(3) & 0.29\\
			Sr(2) & 6g & 0.0513(2) & 0.250 & -0.164(3) & 0.21\\
			B(1) & 6g & 0.704(1) & 0.250 & -0.284(2) & 0.5\\
			B(2) & 2d & 0.152(9) & -0.250 & 0.751(2) & 0.5\\
			O(1) & 6g & 0.540(6) & 0.250 & -0.295 (7) & 0.5\\
			O(2) & 6g & 0.763(4) & 0.024(5) & -0.207(6) & 1.00\\
			O(3) & 6g & -0.027(6) & 0.250 & 0.386(7) & 0.5\\
			O(4) & 6g & 0.708(4) & 0.059(6) & 0.239(6) & 1.00\\
		\end{tabular}
	\end{ruledtabular}
\end{table}

Single crystals of K$_3$YbSi$_2$O$_7$ were grown by the high temperature flux method. A homogeneous mixture of Yb$_2$O$_3$ (Aldrich, 99.9\%) and SiO$_2$ (Aldrich, 99\%) in a molar ratio of 1:12 (for 0.1gm of K$_3$YbSi$_2$O$_7$) and KF flux (2gm, Aldrich, 99.9\%) was transferred to a platinum crucible, covered with a lid. Then, the mixture was sintered at 1000$^{\degree}$C for 12~hrs followed by slow cooling to 800$^{\degree}$C in a cooling rate of 0.5$^{\degree}$C/min. Finally, the furnace was switched off and the sample was furnace cooled to room temperature. Small single crystals were manually removed, washed with distilled water, and dried in air. Single-crystal XRD on a good-quality single crystal was performed at room temperature using a Bruker KAPPA-II diffractometer with a CCD detector and graphite monochromated Mo $K_{\alpha}$ radiation ($\lambda_{\rm avg} \sim 0.71073~\AA$). The data were recorded using APEX3 software and reduced with SAINT/XPREP. An empirical absorption correction was performed using the SADABS program. The crystal structure was solved by direct methods using SHELXT-2018/2 and refined by the full matrix least squares method on $F^2$ using SHELXL2018/3, respectively. The obtained crystal structure is Hexagonal (space group: $P6_3/mmc$) which is consistent with the previous report~\cite{Dabic584}. The details of the refined crystal structure and atomic positions are tabulated in Tables~\ref{Crystal_KYbSiO} and \ref{Positions K3YbSi2O7}, respectively. The anisotropic atomic displacement parameters are given the supplementary material (SM)~\cite{Supplementary}. To further cross-check the phase purity, a large number of single crystals of K$_3$YbSi$_2$O$_7$ were crushed into powder and the powder XRD measurement was performed at room temperature. As shown in Fig.~\ref{Fig2}(b), the Rietveld refinement of the powder XRD pattern confirms high quality phase pure sample. The obtained lattice parameters $a = 5.7323(4)$~\AA, $c = 13.956(3)$~\AA, and $V_{\rm cell} \simeq 397.15$~\AA$^{3}$ agree with the single crystal XRD. Since the crystals were very small, all the measurements on this compound were done on the crushed powder sample.
\begin{table}
	\caption{Details of the structural data of K$_3$YbSi$_2$O$_7$ obtained from the single crystal XRD.}
	\label{Crystal_KYbSiO}
	\begin{tabular}{ccccccc}
		\hline \hline
		\textbf{Crystal data}\\
		Empirical formula & K$_3$YbSi$_2$O$_7$\\
		Formula weight ($M_r$)& 458.52 g \\	
		Crystal system & Hexagonal \\
		Space group & $P6_3/m m c$\\
		$a$ (\AA) & 5.7177(6)\\
		$c$ (\AA) & 13.9132(18)\\
		$V_{\rm cell}$ (\AA$^3$) & 393.91(10)\\       
		$Z$ & 2\\ 
		Calculated crystal density $\rho_{\rm cal}$ & 3.866 mg/mm$^3$\\
		Absorption coefficient ($\mu$) & 13.758 mm$^{-1}$\\ 	
		Crystal size & $0.068 \times 0.048 \times 0.038 $ mm$^3$\\	
		\textbf{Data collection}\\
		Temperature~(K)& 296(2)\\
		Radiation type & Mo$K_{\alpha1}$\\  
		Wavelength ($\lambda$) & 0.71073~\AA\\
		Diffractometer & Bruker KAPPA APEX-II CCD\\	
		$\theta$ range for data collection&2.708$^{\circ}$ to 26.422$^{\circ}$\\
		Index ranges & $-7\leq h\leq 7$,&\\
				&$-7\leq k\leq 7$,&\\
				&$-18\leq l\leq 18$\\
		
			$F$(000) & 422.0\\
				Reflections collected & 2114\\
				Independent reflections & 213 [$R_{\rm int} = 0.0406$] \\
			Data/restraints/parameters & 213/12/19\\
			Final $R$ indexes, $I\geq 2\sigma(I)$&$R_{1}=0.0224$, $\omega R_{2} = 0.0557$& \\	
			Final $R$ indexes, all data & $R_{1}=0.0242$, $\omega R_{2}=0.0570$& \\
			Largest difference peak/hole&1.014 / -0.487 e.\AA$^{-3}$\\
		\textbf{Refinement}\\
		Refinement method&Full-matrix least-squares on $F^2$\\
		Goodness-of-fit on $F^{2}$&1.128\\
		\hline\hline 
	\end{tabular}
\end{table}
\begin{table}
\caption{The Wyckoff positions, atomic coordinates, and isotropic atomic displacement parameters ($U_{\rm iso}$) for K$_3$YbSi$_2$O$_7$, obtained from the single crystal XRD. $U_{\rm iso}$ is defined as one-third of the trace of the orthogonal $U_{\rm ij}$ tensor. The error bars are from the least-square structure refinement. The occupancy is one for all the atoms.}
\label{Positions K3YbSi2O7}
\begin{tabular}{ccccccc}
		\hline \hline
		\multicolumn{1}{p{1cm}}{\centering Atomic \\ sites} & \multicolumn{1}{p{1.3
				cm}}{\centering Wyckoff\\ positions} &\multicolumn{1}{p{2
				cm}}{\centering $x$} &\multicolumn{1}{p{1.3
				cm}}{\centering $y$} &\multicolumn{1}{p{1.3
				cm}}{\centering $z$}  &  $U_{\rm iso}$\\\hline
		K(1) &4f& 1/3 & 2/3 & 0.1085(2) & 0.015(1)\\
		K(2) &2b& 0 & 0 & 1/4 & 0.02(1)\\
        Yb(1) &2a& 0 & 0 & 0 & 0.007 (1)\\           
		O(1)&12k & 0.3642(8) & 0.1821(4) & 0.0841(3) & 0.015(1)\\ 
        O(2) &2d& 2/3 & 1/3 & 1/4 & 0.022(2)\\  
		Si(1)&4f & 2/3 & 1/3 & 0.1367(2) & 0.007(1)\\                            \hline\hline
	\end{tabular}
\end{table}

Magnetization ($M$) as a function of temperature ($T$) was measured in the temperature range 0.4–380~K in different magnetic fields ($H$) using a superconducting quantum interference device (SQUID) magnetometer (MPMS 3, Quantum Design). For achieving temperature below 1.8~K, a $^{3}$He insert (iHelium3) was used in SQUID. The isothermal magnetization ($M$ vs $H$) was measured at different temperatures from 0 to 7~T. The temperature-dependent specific heat [$C_{\rm P}(T)$] at different fields (0~T $\leq\mu_0H\leq$ 9~T) was measured on a small piece of sintered pellet in a large temperature range (0.4~K$\leq T\leq 300$~K) using the standard thermal relaxation technique in a Physical Property Measurement System (PPMS, Quantum Design). A $^{3}$He attachment to the PPMS was used to measure specific heat below 2~K.

%******************************* For ADR **************************************************
%In order to attain lowest temperature below the 1.8~K without using $^{3}$He, we perform ADR using a commercial PPMS. For ADR measurement, we used a 4.02~g sinter (...$^{\degree}$) pellet contains equal amount of silver powder and NaSrYb(BO$_3$)$_2$. Silver powder is used to increase the thermal conduction within the pellet of the insulating NaSrYb(BO$_3$)$_2$. Pellet was kept on the sapphire sample holder. A resistive RuO$_2$ thermometer is glued on the pellet and connected by a very thin manganin wire to minimize heat transfer. This setup is protected from the thermal radiation of the surrounding by using a metallic cap as a shield. In presence of high-vacuum, the sample is precooled to a starting temperature 2~K at a magnetic field ...~T using PPMS. Then the magnetic field is set to zero with a sweep rate ...min$^{-1}$. The sample reaches lowest temperature and warm up back to 2~K due to weak heat flow from the bath. This heat conduction is happened by small amount of exchange gas.
%*********************************************************************************
The crystal electric field (CEF) calculations of both the systems were performed using the point charge model using the PyCrystalField Python package~\cite{Scheie356}. The structural parameters (lattice constants and atomic positions) used for this calculation are taken from the Rietveld refinement (\ref{Refined parameters} and~\ref{Positions K3YbSi2O7}) of the powder XRD data. The six nearest-neighbor oxygen (O$^{2-}$) ligands of the YbO$_6$ octahedra are included in this calculation.

%Magnetic susceptibility was modeled by high-temperature series expansion (HTSE) for spin-1/2 triangular lattice~\cite{Tamura729} as well as full diagonalization (FD) for the 3 × 3 finite lattice with periodic boundary conditions using the ALPS package~\cite{BauerP05001}.

\section{Results}
\subsection{Magnetization} 
\begin{figure*}
\includegraphics[width=0.85\linewidth]{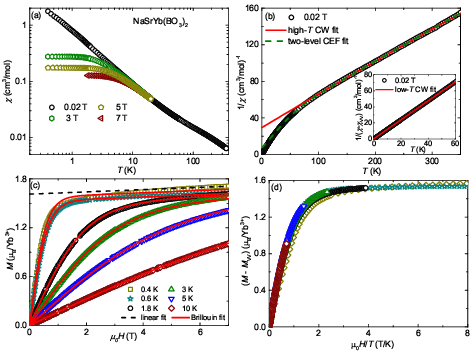}
\caption{(a) $\chi$ vs $T$ of NaSrYb(BO$_3$)$_2$ measured in different applied magnetic fields.
(b) Inverse susceptibility ($1/\chi$) vs $T$ measured at $\mu_0H = 0.02$~T for NaSrYb(BO$_3$)$_2$. The solid and dashed lines represent the high-$T$ CW fit [Eq.~\eqref{CW}] and two-level CEF fit [Eq.~\eqref{CEF_CW}], respectively. Inset: The low-temperature $1/(\chi-\chi_{\rm VV}$) data along with the CW fit. (c) Magnetic isotherms ($M$ vs $H$) of NaSrYb(BO$_3$)$_2$ at different temperatures along with the Brillouin fits. The dashed line represents a linear fit to the high-field data for $T = 0.4$~K. (d) ($M-M_{\rm VV}$) vs $\mu_{0}H/T$ at different temperatures.}
\label{Fig3}
\end{figure*}
\begin{figure*}
\includegraphics[width=0.85\linewidth]{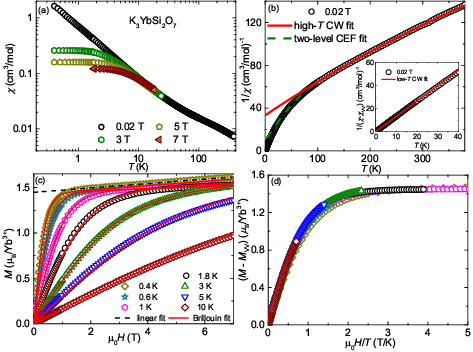}
\caption{(a) $\chi$ vs $T$ of K$_3$YbSi$_2$O$_7$ measured in different applied magnetic fields. (b) Inverse susceptibility ($1/\chi$) vs $T$ measured at $\mu_0H = 0.02$~T for K$_3$YbSi$_2$O$_7$. The solid and dashed lines represent the high-$T$ CW fit [Eq.~\eqref{CW}] and two-level CEF fit [Eq.~\eqref{CEF_CW}], respectively. Inset: The low-temperature $1/(\chi-\chi_{\rm VV})$ data along with the CW fit for $\mu_0H = 0.02$~T. (c) $M$ vs $H$ of K$_3$YbSi$_2$O$_7$ at different temperatures along with the Brillouin fits. The dashed line represents a linear fit to the high-field data at $T = 0.4$~K. (d) ($M-M_{\rm VV}$) vs $\mu_{0}H/T$ at different temperatures.}
\label{Fig4}
\end{figure*}

Temperature-dependent magnetic susceptibility $\chi [\equiv M/H]$ of NaSrYb(BO$_3$)$_2$ and K$_3$YbSi$_2$O$_7$ in different magnetic fields are presented in Fig.~\ref{Fig3}(a) and~\ref{Fig4}(a), respectively. In the high-temperature region, $\chi(T)$ of both compounds follows a typical Curie-Weiss (CW) behaviour. No clear signature of magnetic LRO is observed down to 0.4~K for both compounds. However, a small cusp in $d\chi/dT$ (see SM~\cite{Supplementary}) appears at around $\sim 2.25$~K for NaSrYb(BO$_3$)$_2$ in low magnetic fields, signaling the onset of magnetic LRO, arising possibly due to a tiny amount ($< 1\%$) of Yb$_2$O$_3$ impurity present in the powder sample~\cite{Moon722}. Indeed, it is observed that the majority of the Yb$^{3+}$ based compounds in the polycrystalline form use to have a small amount of Yb$_2$O$_3$ impurity~\cite{Tokiwa42,Arjun224415}.
%Measurements of $\chi(T)$ at 200~Oe using both zero-field-cooled (ZFC) and field-cooled (FC) protocols show no bifurcation (not shown here) down to 1.8~K, which rules out the spin-glass (SG) transition or spin freezing at low temperature. 
The inverse magnetic susceptibility [$1/\chi(T)$], in the high-temperature region, was well fitted by the modified CW law
\begin{equation}\label{CW}
\chi(T) = \chi_0 + \frac{C}{T - \theta_{\rm CW}}.
\end{equation}
Here, $\chi_0$ is the combination of $T$-independent core diamagnetic ($\chi_{\rm dia}$) and Van-Vleck paramagnetic ($\chi_{\rm VV}$) susceptibilities. In the second term of Eq.~\eqref{CW}, $C$ is the Curie constant and $\theta_{\rm CW}$ is the CW temperature.

The CW fit for $T \ge 90$~K yields the parameters $\chi_0^{\rm HT} \simeq 3.1 \times 10^{-5}$~cm$^3$/mol, $C^{\rm HT} \simeq 2.65$~cm$^3$K/mol, and $\theta_{\rm CW}^{\rm HT} \simeq -78.8$~K for NaSrYb(BO$_3)_2$ [see Fig.~\ref{Fig3}(b)]. Similarly, for K$_3$YbSi$_2$O$_7$, the CW fit for $T \ge 100$~K returns $\chi_0^{\rm HT} \simeq 1.7 \times 10^{-3}$~cm$^3$/mol, $C^{\rm HT} \simeq 2.67$~cm$^3$K/mol, and $\theta_{\rm CW}^{\rm HT} \simeq -94$~K [see Fig.~\ref{Fig4}(b)]. From the $C^{\rm HT}$ value, the effective moment $\mu_{\rm eff}^{\rm HT}$ $[=\sqrt{\left(\frac{3k_{\rm B}C}{N_{\rm A}}\right)}\mu_{\rm B}$, where $N_{\rm A}$ is the Avogadro’s number, $\mu_{\rm B}$ is the Bohr magneton, and $k_{\rm B}$ is the Boltzmann constant] is calculated to be $\sim 4.59~\mu_{\rm B}$ and $\sim 4.62$~$\mu_{\rm B}$ for NaSrYb(BO$_3)_2$ and K$_3$YbSi$_2$O$_7$, respectively. These values of $\mu_{\rm eff}^{\rm HT}$ are close to the expected value $\sim 4.54~\mu_{\rm B}$ for a free Yb$^{3+}$ ion ($J = 7/2$ and Land\'e $g$-factor $g = 1.14$) with $4f^{13}$ electronic configuration.
Here, the large negative value of $\theta_{\rm CW}^{\rm HT}$ does not imply the presence of strong AFM interactions between the Yb$^{3+}$ ions. Rather, it indicates the effect of CEF excitations at high temperatures. For the Yb$^{3+}$ ($J=7/2$) ion, one expects the spin-orbit entangled eight-fold degenerate ground state to split into four Kramer's doublets (with quantum numbers $J_z = \pm \frac{1}{2}$, $\pm \frac{3}{2}$, $\pm \frac{5}{2}$, and $\pm \frac{7}{2}$) due to CEF interaction. At high temperatures, all the higher energy doublets get thermally populated and contribute to $\theta_{\rm CW}$. However, when the temperature is lowered below the energy gap between the ground state and first excited state doublets, only the lowest energy doublet is populated and contributes to the ground state properties. In the Yb$^{3+}$ based compounds, the lowest  Kramers' doublet with effective $J_{\rm eff} = 1/2$ typically controls magnetic properties at low temperatures, while the higher-lying doublets produce a sizable Van-Vleck contribution ($\chi_{\rm VV}$)~\cite{Guchhait144434,Somesh064421,Ranjith115804}.

As observed in Figs.~\ref{Fig3}(b) and \ref{Fig4}(b), $1/\chi(T)$ displays a clear slope change below about 50~K.
$1/\chi(T)$ after correcting the Van-Vleck contribution [i.e. $1/(\chi-\chi_{\rm VV})$] shows a distinct linear regime in the low-temperature side [see insets of Figs.~\ref{Fig3}(b) and \ref{Fig4}(b)]. Here, $\chi_{\rm VV}$ was obtained from the analysis of magnetization isotherm at $T = 0.4$~K (discussed later). A CW fit in the temperature range 2 to 20~K yields ($C^{\rm LT}\simeq 0.84$~cm$^3$K/mol and $\theta_{\rm CW}^{\rm LT} \simeq -0.1$~K) for NaSrYb(BO$_3$)$_2$ and ($C^{\rm LT} \simeq 0.74$~cm$^3$K/mol and $\theta_{\rm CW}^{\rm LT} \simeq -0.1$~K) for K$_3$YbSi$_2$O$_7$, respectively. The small negative value of $\theta_{\rm CW}^{\rm LT}$ suggests a very weak and dominant antiferromagnetic (AFM) interaction among the Yb$^{3+}$ ($J_{\rm eff} = 1/2$) ions.
%According to the mean-field approximation, $\theta_{\rm CW}$ is the sum of all possible exchange couplings present in a system~\cite{Kittel1986}. In these systems, it is also possible that the presence competing ferromagnetic (FM) and AFM interactions cancel each other, resulting in very small values of $\theta_{\rm CW}^{\rm LT}$.
The obtained $C^{\rm LT} \simeq 0.84$~cm$^3$K/mol and $0.74$~cm$^3$K/mol values are equivalent to an effective moment of $\mu_{\rm eff}^{\rm LT} \simeq~2.6\mu_{\rm B}$ and $\mu_{\rm eff}^{\rm LT} \simeq~2.45\mu_{\rm B}$ for NaSrYb(BO$_3$)$_2$ and K$_3$YbSi$_2$O$_7$, respectively. These values of $\mu_{\rm eff}^{\rm LT}$ are reminiscent of an effective spin $J_{\rm eff} = 1/2$ ground state with an average $g_{\rm ave} \simeq 3$ for NaSrYb(BO$_3$)$_2$ and $g_{\rm ave} \simeq 2.8$ for K$_3$YbSi$_2$O$_7$. The reduced value of the effective moment or $J_{\rm eff} = 1/2$ at low-$T$s can be attributed to the effect of CEF splitting and depopulation of higher energy doublets. These $g$-values are in close agreement with the CEF calculations (discussed later) as well as the electron spin resonance (ESR) experiments on other Yb$^{3+}$ based systems~\cite{Ranjith180401,Ranjith224417}.

In order to have a rough estimation of CEF energy gap, we also fitted $1/\chi(T)$ by a simplified effective two-level model~\cite{Mugiraneza95}
\begin{equation}\label{CEF_CW}
	\chi(T) = \chi_0 + \frac{1}{8(T - \theta_{\rm CW})}\times\left[\frac{\mu_{\rm eff,0}^{2}+\mu_{\rm eff,1}^{2}e^{-\left(\frac{\Delta^{\rm CEF}}{k_{\rm B}T}\right)}}{1+e^{-\left(\frac{\Delta^{\rm CEF}}{k_{\rm B}T}\right)}}\right].
\end{equation}
Here, $\Delta^{\rm CEF}$ is the energy gap between the ground state and the first excited CEF doublets. $\mu_{\rm eff,0}$ and $\mu_{\rm eff,1}$ are the effective moments of the ground state and the first excited CEF levels, respectively. The two-level CEF fit for $T\geq 25$~K regime yields $\chi_0^{\rm CEF} \simeq 2.45 \times 10^{-4}$~cm$^3$/mol, $\mu_{\rm eff,0}^{\rm CEF} \simeq 3.33\mu_{\rm B}$/Yb$^{3+}$, $\mu_{\rm eff,1}^{\rm CEF} \simeq 5.53~\mu_{\rm B}$/Yb$^{3+}$, $\Delta^{\rm CEF}/k_{\rm B} \simeq 
 215$~K, and $\theta_{\rm CW}^{\rm CEF} \simeq 
-10.13$~K. Similarly, in the case of K$_3$YbSi$_2$O$_7$, this fit for $T\geq 25$~K gives $\chi_0^{\rm CEF} \simeq 1.6 \times 10^{-3}$~cm$^3$/mol, $\mu_{\rm eff,0}^{\rm CEF}\simeq 3.3 \mu_{\rm B}$/Yb$^{3+}$, $\mu_{\rm eff,1}^{\rm CEF}\simeq 5.56~\mu_{\rm B}$/Yb$^{3+}$, $\Delta^{\rm CEF}/k_{\rm B} \simeq 229$~K, and $\theta_{\rm CW}^{\rm CEF}\simeq -13.7$~K. These values of $\Delta^{\rm CEF}/k_{\rm B}$ are in the same order of magnitude as that of our CEF calculations, presented later. However, the obtained $\theta_{\rm CW}^{\rm CEF}$ values deviate significantly from the ones derived from the low-$T$ $\chi(T)$ analysis~\cite{Arjun014013,Arjun224415,Pula014412}. As these parameters solely depend on the symmetry and strength of the crystal field environment, a four-level fit may be required. However, a large number of fitting parameters using a four-level fit often result unreliable values.

The magnetization isotherms $M(H)$ of NaSrYb(BO$_3$)$_2$ and K$_3$YbSi$_2$O$_7$ measured up to 7~T at different temperatures are shown in Figs.~\ref{Fig3}(c) and \ref{Fig4}(c), respectively. The $M(H)$ curve at $T = 0.4$~K shows the saturation of magnetization at around $\mu_0H_{\rm S}\simeq 1$~T for both compounds and then increases linearly in higher fields due to the Van-Vleck paramagnetic contribution. The saturation magnetization $M_{\rm sat}$ and $\chi_{\rm VV}$ of both compounds are obtained from the slope and $y$-intercept of a straight line fit to the data above 5~T. We obtained $\chi_{\rm VV}\simeq 7.04\times10^{-3}$ cm$^{3}$/mol and $M_{\rm sat} \simeq 1.6$~$\mu_{\rm B}$ for NaSrYb(BO$_3$)$_2$ and $\chi_{\rm VV}\simeq 7.66 \times 10^{-3}$~cm$^{3}$/mol and $M_{\rm sat} \simeq 1.45$~$\mu_{\rm B}$ for K$_3$YbSi$_2$O$_7$, respectively. The value of $M_{\rm sat}$ ($M_{\rm sat} = g_{\rm ave}J_{\rm eff}\mu_{\rm B}$) corresponds to $J_{\rm eff} =1/2$ with an average value of $g_{\rm ave} \simeq 3.2$ and 2.9 for NaSrYb(BO$_3$)$_2$ and K$_3$YbSi$_2$O$_7$, respectively. These values of $g_{\rm ave}$ are close to the obtained $g$-value for other Yb$^{3+}$-based compounds at low-temperatures~\cite{Ranjith224417,Li167203}.

As we noticed from the $\chi(T)$ analysis that the value of $\theta_{\rm CW}^{\rm LT}$ is negligibly small, reflecting very weak magnetic correlations at low temperatures for both the compounds. Therefore, we tried to model the magnetic isotherms at different temperatures by the following expression~\cite{Thamban255801,Biswal134420}
\begin{equation}
	\label{MH imp}
	M(H)=\chi_{\rm VV}H + N_{\rm A}g\mu_{\rm B}J_{\rm eff}BJ_{\rm eff}(x),
\end{equation}
assuming uncorrelated spins at low temperatures.
Here, $BJ_{\rm eff}(x)$ is the Brillouin function and $x= g\mu_{\rm B}J_{\rm eff}H/(k_{\rm B}T)$. For $J_{\rm eff} = 1/2$, the Brillouin function reduces to $BJ_{\rm eff}(x)= \text{tanh}(x)$~\cite{Kittel1986}. While fitting the $M(H)$ curves by Eq.~\eqref{MH imp} we fixed the value of $\chi_{\rm VV}$. All the curves of both compounds at low temperature are well fitted by Eq.~\eqref{MH imp}, which comprehends the uncorrelated paramagnetic spins at low temperatures. In order to further visualize this behaviour, we plotted the Van-Vleck subtracted magnetization, $M-M_{\rm VV}$ vs $\mu_0H/T$ at different temperatures as shown in Figs.~\ref{Fig3}(d) and~\ref{Fig4}(d). All the curves almost collapse onto a single curve, which is a clear indication of the paramagnetic nature of the spins and very weak magnetic correlation at low temperatures.

\subsection{Specific Heat}
\begin{figure*}
	\includegraphics[width=\linewidth]{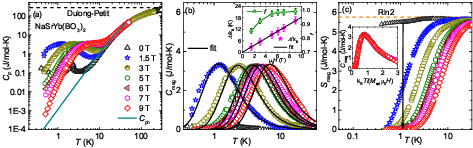}
	\caption{(a) Variation of $C_{\rm p}$ with temperature for NaSrYb(BO$_3$)$_2$ measured in different applied magnetic fields. The dark cyan solid line represents the phonon specific heat ($C_{\rm Ph}$) obtained by the polynomial fit. The horizontal dashed line indicates the expected Dulong-Petit value for this compound. (b) $C_{\rm mag}$ vs $T$ at different magnetic fields along with two-level Schottky fits. Inset: $\Delta/k_{\rm B}$ vs $\mu_0H$ (left $y$-axes) and $f$ vs $\mu_0H$ (right $y$-axes). The solid line is the linear fit to $\Delta/k_{\rm B}(H)$. (c) The magnetic entropy change, $S_{\rm mag}$ vs $T$ at different applied fields. The zero-field entropy is scaled to match $Rln2$ and the vertically downward arrow indicates the entropy change between 0 and 9~T. Inset: Scaling $C_{\rm mag}$ vs $k_{\rm B}T/(M_{\rm sat}\mu_0H)$ plots in different magnetic fields.}
	\label{Fig5}
\end{figure*}
\begin{figure*}
\includegraphics[width=\linewidth]{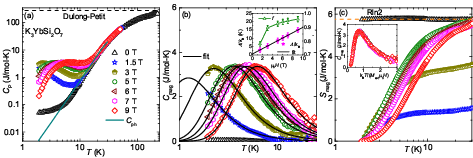}
\caption{(a) Variation of $C_{\rm p}$ with temperature for K$_3$YbSi$_2$O$_7$ measured in different applied magnetic fields. The dark cyan solid line represents the phonon specific heat ($C_{\rm Ph}$) obtained by the polynomial fit. The horizontal dashed line indicates the expected Dulong-Petit value of this compound. (b) $C_{\rm mag}$ vs $T$ at different magnetic fields along with two-level Schottky fits. Inset: $\Delta/k_{\rm B}$ vs $\mu_0H$ (left $y$-axes) and $f$ vs $\mu_0H$ (right $y$-axes). The solid line is the linear fit of $\Delta/k_{\rm B}(H)$. (c) The magnetic entropy change, $S_{\rm mag}$ vs $T$ at different applied fields. The zero-field entropy is scaled to match $R ln 2$. Inset: Scaling $C_{\rm mag}$ vs $k_{\rm B}T/(M_{\rm sat}\mu_0H)$ plots in different magnetic fields.}
\label{Fig6}
\end{figure*}
Temperature-dependent specific heat [$C_{\rm P}(T)$] of NaSrYb(BO$_3$)$_2$ and K$_3$YbSi$_2$O$_7$ measured down to 0.4~K in different applied fields is shown in Figs.~\ref{Fig5}(a) and~\ref{Fig6}(a), respectively. Similar to $\chi(T)$, no magnetic LRO could be detected down to 0.4~K for both the compounds. In Fig.~\ref{Fig5}(a), we observed a sharp $\lambda$-type anomaly around 2.23~K in zero field which corresponds to the magnetic ordering due to a small fraction (see SM~\cite{Supplementary}) of extrinsic Yb$_2$O$_3$ phase in NaSrYb(BO$_3$)$_2$.
%\textbf{The calculated entropy (see SM) associated with this impurity peak yields a negligible fraction ($\sim 1$~\%) of $Rln2$.}
In a magnetic insulator, the total specific heat $C_{\rm P}(T)$ is the sum of phonon/lattice contribution [$C_{\rm ph}(T)$], which dominates in the high-temperature region and magnetic specific heat [$C_{\rm mag}(T)$], which dominates in the low-temperature region. In order to extract $C_{\rm mag}(T)$, we subtracted $C_{\rm ph}(T)$ from $C_{\rm P}(T)$. To quantify $C_{\rm ph}(T)$, the zero-field $C_{\rm P}(T)$ was fitted by a polynomial ($=aT^{3}+bT^{5}+cT^{7}$) function for $T \ge 10$~K~\cite{Nath054409}. The obtained values of fitting parameters ($a$, $b$, and $c$) are given in SM~\cite{Supplementary}. The fit was extrapolated down to 0.4~K and then subtracted from the total specific heat. The obtained $C_{\rm mag}(T)$ data at various fields are plotted in Figs.~\ref{Fig5}(b) and~\ref{Fig6}(b) for NaSrYb(BO$_3$)$_2$ and K$_3$YbSi$_2$O$_7$, respectively. The magnetic entropy [$S_{\rm mag}(T)$] released at different fields are estimated by integrating $C_{\rm mag}/T$ over temperature. The estimated $S_{\rm mag}(T)$ at different fields are presented in Figs.~\ref{Fig5}(c) and~\ref{Fig6}(c) for NaSrYb(BO$_3$)$_2$ and K$_3$YbSi$_2$O$_7$, respectively. For both the compounds, $S_{\rm mag}(T)$ attains a value $\sim 5.6$~J.mol$^{-1}$.K$^{-1}$ at around 30~K in high fields, which is close to the expected value for a two-level system ($R ln2 = 5.76$~J.mol$^{-1}$.K$^{-1}$, where, $R$ is the universal gas constant). This confirms a Kramers' doublet with $J_{\rm eff} = 1/2$ ground state  at low temperatures for both the compounds.

In zero-field, $C_{\rm P}(T)$ shows an upturn towards lower temperatures, implying the development of AFM correlation and the entropy accumulation associated with the lowest Kramers' doublet of Yb$^{3+}$. In zero-field, the entire entropy of $Rln2$ associated with the lowest Kramers' doublet could not be recovered as we did not have data below 0.4~K. Therefore, the zero-field entropy in Figs.~\ref{Fig5}(c) and~\ref{Fig6}(c) is vertically shifted to match with the high-field data. The ground-state doublet is split by the applied field into $J_z = +1/2$ and $J_z = -1/2$ levels, causing a Schottky anomaly (broad maximum in specific heat), which shifts toward higher temperatures with increasing field. To estimate the Schottky contribution, $C_{\rm mag}(T)$ was fitted using the two-level Schottky function
\begin{equation}\label{Schottky}
C_{\rm Sch} (T, H) = fR\left(\frac{\Delta}{k_{B}T}\right)^2\frac{e^{\left(\frac{\Delta}{k_{B}T}\right)}}{\left[e^{\left(\frac{\Delta}{k_{B}T}\right)}+1\right]^{2}}.
\end{equation}
Here, $f$ is the molar fraction of free spins and $\Delta/k_{\rm B}$ is the crystal field energy gap between the Zeeman levels of split ground state doublet. As shown in Figs.~\ref{Fig5}(b) and~\ref{Fig6}(b), $C_{\rm mag}(T)$ data of both compounds are well fitted by Eq.~\eqref{Schottky}. The estimated fitting parameters, $f$ and $\Delta/k_{\rm B}$ are plotted as a function of applied field in the inset of Figs.~\ref{Fig5}(b) and~\ref{Fig6}(b) for NaSrYb(BO$_3$)$_2$ and K$_3$YbSi$_2$O$_7$, respectively. The value of $f$ increases with $H$ and almost saturates to 1 in higher fields. This indicates that the applied field causes splitting of the energy levels and excites the free Yb$^{3+}$ spins to higher energy states. Below the saturation field, a small fraction of spins are correlated, which decreases with increasing field. Above the saturation field all the free spins are excited and $S_{\rm mag}$ saturates to a value $Rln2$. $\Delta/k_{\rm B}$ increases linearly with $H$ and a straight line fit yields a very small zero-field energy gap $\Delta/k_{\rm B}(0) \simeq 0.035$~K and $\sim 0.5$~K, respectively. The small value of $\Delta/k_{\rm B}(0)$ suggest the presence of a weak intrinsic field in these systems~\cite{Kundu117206}. Using the value of $\Delta/k_{\rm B} \simeq 16.13$~K at 9~T, $g~(=\Delta/\mu_{\rm B}H)$ is calculated to be $\sim 2.75$ for NaSrYb(BO$_3$)$_2$. Similarly, using $\Delta/k_{\rm B} \simeq 14.98$~K at 9~T, we obtained $g \simeq 2.5$. These $g$ values for both the compounds are consistent with other Yb$^{3+}$ based systems~\cite{Ranjith224417}.

A scaling plot of $C_{\rm mag}$ vs $k_{\rm B}T/M_{\rm sat}\mu_0H$ for both the compounds are shown in the inset of Figs.~\ref{Fig5}(c) and~\ref{Fig6}(c). The scaling of all the fields for $\mu_0H \geq 0.5$~T collapse in to a single curve which indicates that the interaction strength between the Yb$^{3+}$ ions is very small, which is consistent with very small value of $\theta_{\rm CW}^{\rm LT}$. Further, the Kramers' doublets can be characterized by a dimensionless quantity $R = \left(\frac{\mu_{\rm eff}}{M_{\rm sat}}\right)^{2} = 3$. Using the experimental values of $\mu_{\rm eff}^{\rm LT}$ and $M_{\rm sat}$, the value of $R$ is calculated to be $R \simeq 3.04$ and 2.9 for NaSrYb(BO$_3$)$_2$ and K$_3$YbSi$_2$O$_7$, respectively, which are close to the theoretical expected value for a $J_{\rm eff} = 1/2$ system~\cite{Guo094404,Guchhait144434}.

%\subsection{Adiabatic demagnetization refrigeration}
%The ADR cooling performance of NaSrYb(BO$_3$)$_2$ using PPMS is shown in Fig.~. Sample temperature is plotted in left $y$-axis whereas applied magnetic field is plotted at right-$y$-axis. We achived ..~K final temperature by the ADR measurement. The warm-up time for this compound is ... . From the warm-up curve, one can estimate the specific heat of the material using the following equation
%\begin{equation}\label{Cp_ADR}
% \.{Q} = C_{\rm ADR}\.{T}.
%\end{equation}
%Here $\.{Q}$ is the constant heat input per unit time, $C_{\rm ADR}$ is the magnetic specific heat of the ADR pellet, and $\.{T}$ is the time derivative of temperature during warming. The direct measured magnetic specific heat using ADR is in close agreement with $C_{\rm ADR}$ for $\.{Q}=..$~$\mu$W.
%********************************ADR*****************

\subsection{CEF analysis}
To understand the single-ion effects on the ground state properties and to estimate the CEF energy level scheme of the doublets, we performed the CEF calculation using the point charge approximation method~\cite{Newman_Ng_2000}. Yb$^{3+}$ has a 4$f^{13}$ electronic configuration ($L = 3$, $S = 1/2$, and $J = 7/2$). This spin-orbit coupled eight-fold ($2J+1$) degenerate states further split into four Kramer's doublets with quantum numbers $J_z = \pm \frac{1}{2}$, $\pm \frac{3}{2}$, $\pm \frac{5}{2}$, and $\pm \frac{7}{2}$. According to the Stevens convention, the CEF Hamiltonian can be written as~\cite{Stevens209}
\begin{equation}\label{CEF}
\mathcal{H}_{\rm CEF} = \sum_{l,m}B_l^m\hat{O}_l^m.
\end{equation}
Here, $\hat{O}^m_l$ are the Stevens operators, which are related to the angular momentum operators~\cite{Huthchings227,Stevens209}. $B^m_l$ are the multiplicative factors, called CEF parameters, which are related to the electronic structure of the rare-earth materials. Here, the even integer $l$ varies from 0 to 6 for $f$ electrons and the integer $m$ ranges from $-l$ to $l$. In NaSrYb(BO$_3$)$_2$, distorted YbO$_6$ octahedra generates a low-symmetry CEF environment ($C_{2h}$) around the Yb$^{3+}$ ions. On the other hand, the regular YbO$_6$ octahedra of K$_3$YbSi$_2$O$_7$ produces a symmetric CEF environment ($D_{2h}$) around the Yb$^{3+}$ ions.

\begin{figure}
\includegraphics[scale=0.4]{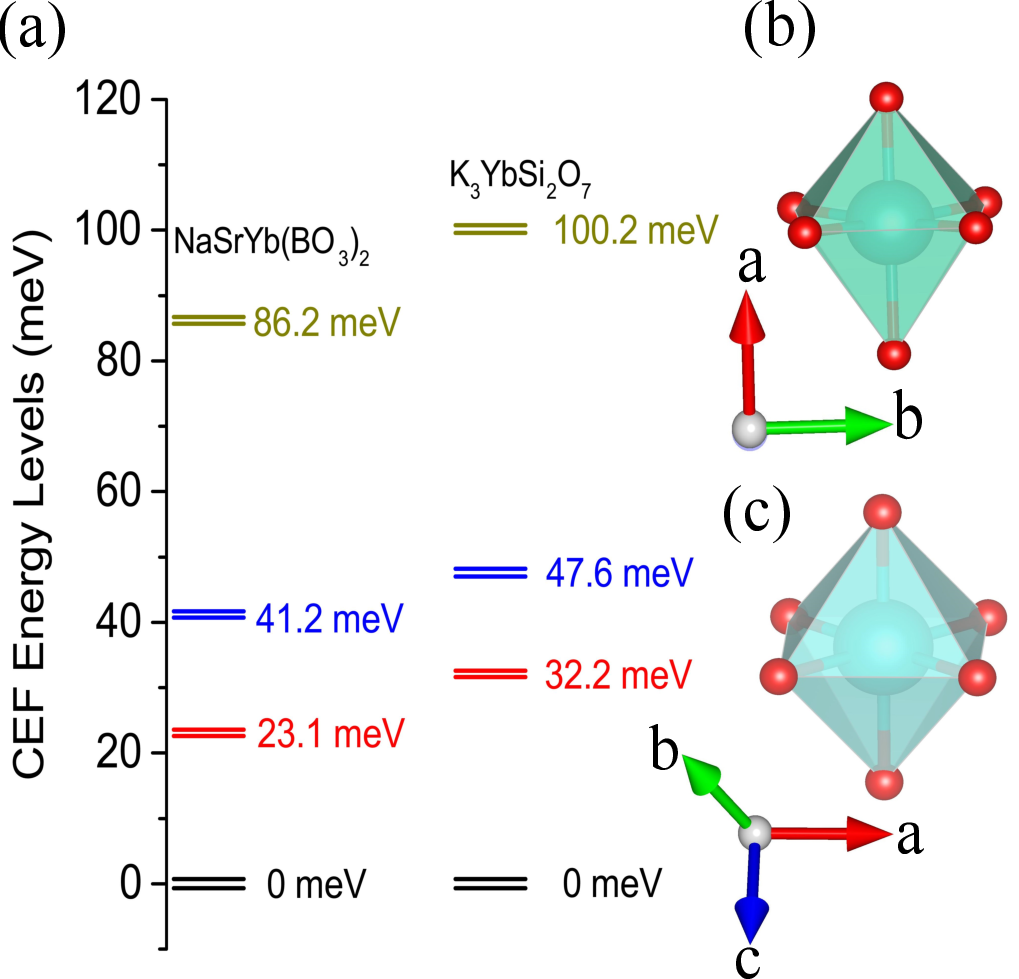}
\caption{(a) Schematic representation of CEF energy levels of NaSrYb(BO$_3$)$_2$ (0, 23.1, 41.2, and 86.2~meV) and K$_3$YbSi$_2$O$_7$ (0, 32.2, 47.6, and 100.2 meV) obtained from the point charge calculations. (b) Distorted YbO$_6$ octahedra of NaSrYb(BO$_3$)$_2$ (c) Regular YbO$_6$ octahedra of K$_3$YbSi$_2$O$_7$ formed by Yb$^{3+}$ and O$^{2-}$ ions that generate CEF.}
\label{Fig7}
\end{figure}

\begin{table}[h!]
   \centering
	\setlength{\tabcolsep}{0.2cm}
	\caption{Calculated CEF parameters for NaSrYb(BO$_3$)$_2$ and K$_3$YbSi$_2$O$_7$.}
	\label{CEF_Para}
	\begin{tabular}{|c| c |c| c |}
		\hline \hline
        \multicolumn{2}{|c|}{NaSrYb(BO$_3$)$_2$}&\multicolumn{2}{|c|}{K$_3$YbSi$_2$O$_7$}\\
        \hline
		$B_l^m$ (meV) & Values & $B_l^m$ (meV) & Values\\\hline 
		$B_2^0$ & $-1.233\times10^{-1}$ & $B_2^0$ & $9.252\times10^{-1}$\\
        $B_2^1$ & -1.789 & $B_2^1$ & $1\times10^{-8}$\\
        $B_2^2$ &  1.6176 & $B_4^0$ & $3.454\times10^{-1}$\\ 
		$B_4^0$ & $-4.032\times10^{-2}$ & $B_4^3$ & $-9.055\times10^{-1}$\\           
		$B_4^1$ &  $5.369\times10^{-3}$ & $B_6^0$ & $1.595\times10^{-4}$\\
        $B_4^2$ &  $2.482\times10^{-3}$ & $B_6^3$ & $3.768\times10^{-3}$ \\           
		$B_4^3$ & $-2.675\times10^{-2}$ & $B_6^6$ & $1.782\times10^{-3}$\\
        $B_4^4$ &  $2.026\times10^{-1}$ & &\\ 
        $B_6^0$ &  $5.911\times10^{-5}$ & &\\
        $B_6^1$ &  $-4.117\times10^{-4}$ & &\\
        $B_6^2$ &  $3.352\times10^{-4}$ & &\\
        $B_6^3$ &  $-3.782\times10^{-4}$ & &\\
        $B_6^4$ &  $1.909\times10^{-3}$ & &\\
        $B_6^5$ &  $2.795\times10^{-4}$ & &\\
        $B_6^6$ &  $-6.849\times10^{-4}$ & &\\
    \hline \hline
	\end{tabular}
\end{table}
%\begin{table}[h!]
%   \centering
%	\setlength{\tabcolsep}{0.4cm}
%	\caption{Calculated CEF parameters for K$_3$YbSi$_2$O$_7$}
%	\label{CEF1_KYbSiO}
%	\begin{tabular}{|c | l | l|}
%		\hline \hline
%		$B_l^m$ (meV) & Values  \\\hline 
%		$B_2^0$ & $9.252\times10^{-1}$ \\
%		$B_2^1$ & $1\times10^{-8}$\\            
%		$B_4^0$ & $3.454\times10^{-1}$\\                      
%       $B_4^3$ & $9.055\times10^{-1}$\\
%       $B_6^0$ & $1.595\times10^{-4}$\\
%       $B_6^3$ & $3.768\times10^{-3}$ \\
%       $B_6^6$ & $1.782\times10^{-3}$\\
%    \hline \hline
%	\end{tabular}
%\end{table}
The allowed CEF parameters calculated for these two compounds using the point charge model are tabulated in Table~\ref{CEF_Para}.
%and~\ref{CEF1_KYbSiO} for NaSrYb(BO$_3$)$_2$ and K$_3$YbSi$_2$O$_7$, respectively. 
These parameters determine the actual CEF Hamiltonian of these compounds [Eq.~\eqref{CEF}]. Next, we diagonalized the Hamiltonian and obtained the CEF energy eigenvalues of these compounds. The obtained energy eigenvalues are 0, 23.13, 41.2, and 86.2~meV, corresponding to four Kramers' doublets of NaSrYb(BO$_3$)$_2$, as shown in Fig.~\ref{Fig7}. Similarly, the obtained CEF energy eigenvalues of K$_3$YbSi$_2$O$_7$ are 0, 32.2, 47.6, and 100.2~meV. From the CEF Hamiltonian [Eq.~\eqref{CEF}], the wave functions corresponding to all the Kramers' doublets can be written as
\begin{equation}\label{CEF_wave_vector}
   |\psi_k,\pm\rangle=\sum_{m_J=-\frac{7}{2}}^{m_J=\frac{7}{2}}C_{m_J}^{k,\pm}\left|J=\frac{7}{2},m_J \right\rangle.
\end{equation}
Here, $C_{m_J}^{k,\pm}$ are the weighted coefficients of the eigenstates and $k$ = 0, 1, 2, and 3 represent the CEF energy levels. The full list of energy eigenvalues and the corresponding coefficients ($C_{m_J}^{k,\pm}$) of different eigenstates for NaSrYb(BO$_3$)$_2$ are listed in Table~\ref{Eigenvalue_and_Eigervector}. The wave function of the ground state doublet (lowest energy doublet) of NaSrYb(BO$_3$)$_2$ is obtained to be
\begin{align}\label{Wavefunction}
\begin{split}
|\psi_0,\pm\rangle = \pm0.734\left|\pm\frac{1}{2}\right\rangle -0.0038\left|\mp\frac{1}{2}\right\rangle \pm0.027\left|\pm\frac{3}{2}\right\rangle &\\ \mp0.0472\left|\mp\frac{3}{2}\right\rangle \mp0.103\left|\pm\frac{5}{2}\right\rangle \pm0.058\left|\mp\frac{5}{2}\right\rangle \mp0.669\left|\mp\frac{7}{2}\right\rangle.
\end{split}
\end{align}
Similarly, the wave function of lowest-energy doublet of K$_3$YbSi$_2$O$_7$ is
\begin{equation}
\label{gswvfunc_KYbSiO}
{|\psi_{0},\pm\rangle} = 0.573\biggr|\pm\frac{1}{2}\biggr> \pm 0.784\biggr|\pm\frac{5}{2}\biggr> \pm 0.238\biggr|\pm\frac{7}{2}\biggr>.
\end{equation}
The complete list of energy eigenvalues and the corresponding coefficients ($C_{m_J}^{k,\pm}$) of all the doublets of K$_3$YbSi$_2$O$_7$ are tabulated in Table~\ref{Eigenvalue_and_Eigervector1}.

\begin{table*}
\caption{Energy eigenvalues and the coefficients ($C_{m_J}^{k,\pm}$) corresponding to different eigenstates of the CEF Hamiltonian for NaSrYb(BO$_3$)$_2$.}
\label{Eigenvalue_and_Eigervector}
\begin{ruledtabular}
\begin{tabular}{c|cccccccccc}
$E$ (meV) & $|-\frac{7}{2}\rangle$ & $|-\frac{5}{2}\rangle$ & $|-\frac{3}{2}\rangle$ & $|-\frac{1}{2}\rangle$ & $|\frac{1}{2}\rangle$ & $|\frac{3}{2}\rangle$ & $|\frac{5}{2}\rangle$ & $|\frac{7}{2}\rangle$ \tabularnewline
 \hline 
0.00 & -0.6697 & 0.0576 & -0.0472 & -0.0038 & 0.7336 & 0.0276 & -0.103 & 0 \tabularnewline
0.00 & 0 & 0.103 & 0.0276 & -0.7336 & -0.0038 & 0.0472 & 0.0576 & 0.6697 \tabularnewline
23.13 & 0.216 & -0.202 & 0.001 & -0.469 & 0.18 & 0.618 & -0.038 & -0.52 \tabularnewline
23.13 & -0.52 & 0.038 & 0.618 & -0.18 & -0.469 & -0.001 & -0.202 & -0.216 \tabularnewline
41.2 & -0.182 & 0.499 & -0.493 & -0.328 & -0.19 & -0.31 & 0.247 & -0.418 \tabularnewline
41.2 & -0.418 & -0.247 & -0.31 & -0.19 & -0.328 & 0.493 & 0.499 & 0.182 \tabularnewline
86.2 & 0 & 0.556 & -0.224 & 0.199 & -0.155 & 0.477 & -0.57 & 0.16  \tabularnewline
86.2 & 0.16 & 0.57 & 0.477 & 0.155 & 0.199 & 0.224 & 0.556 & 0 \tabularnewline
\end{tabular}\end{ruledtabular}
\end{table*}
\begin{table*}
\caption{Energy eigenvalues and the coefficients ($C_{m_J}^{k,\pm}$) corresponding to different eigenstates of the CEF Hamiltonian for K$_3$YbSi$_2$O$_7$.}
\label{Eigenvalue_and_Eigervector1}
\begin{ruledtabular}
\begin{tabular}{c|cccccccccc}
$E$ (meV) & $|-\frac{7}{2}\rangle$ & $|-\frac{5}{2}\rangle$ & $|-\frac{3}{2}\rangle$ & $|-\frac{1}{2}\rangle$ & $|\frac{1}{2}\rangle$ & $|\frac{3}{2}\rangle$ & $|\frac{5}{2}\rangle$ & $|\frac{7}{2}\rangle$ \tabularnewline
 \hline 
0.00 & 0 & -0.784 & 0 & 0 & 0.573 & 0 & 0 & 0.238 \tabularnewline
0.00 & -0.238 & 0 & 0 & 0.573 & 0 & 0 & 0.784 & 0 \tabularnewline
32.2 & 0 & 0 & 0.999 & 0 & 0 & 0.052 & 0 & 0 \tabularnewline
32.2 & 0 & 0 & 0.052 & 0 & 0 & -0.999 & 0 & 0 \tabularnewline
47.6 & 0.541 & 0 & 0 & -0.592 & 0 & 0 & 0.597 & 0 \tabularnewline
47.6 & 0 & -0.597 & 0 & 0 & -0.592 & 0 & 0 & -0.541 \tabularnewline
100.2 & 0 & 0.169 & 0 & 0 & 0.566 & 0 & 0 & -0.807  \tabularnewline
100.2 & -0.807 & 0 & 0 & -0.566 & 0 & 0 & 0.169 & 0 \tabularnewline
\end{tabular}\end{ruledtabular}
\end{table*}

\begin{figure*}
\includegraphics[width=\linewidth]{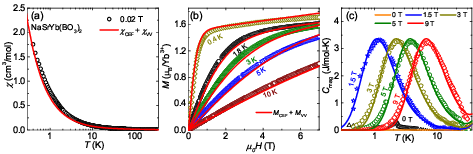}
\caption{ (a) Experimental $\chi(T)$ of NaSrYb(BO$_3$)$_2$ is compared with the calculated susceptibility using the CEF model. (b) Simulated isothermal magnetization at different temperatures using the CEF model and compared with the experimental data. (b) Calculation of CEF specific heat ($C_{\rm CEF}$ vs $T$) in different magnetic fields compared with the experimental $C_{\rm mag}$.}
\label{Fig8}
\end{figure*}
\begin{figure*}
\includegraphics[width=\linewidth]{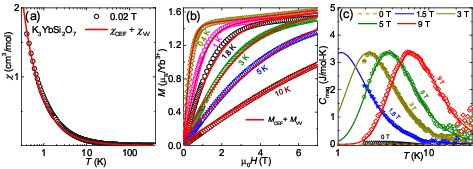}
\caption{ (a) $\chi(T)$ of K$_3$YbSi$_2$O$_7$ is compared with the calculated susceptibility data using the CEF model. (b) Simulated isothermal magnetization at different temperatures using the CEF model and compared with the experimental data. (b) Calculation of CEF specific heat ($C_{\rm CEF}$ vs $T$) in different magnetic fields compared with the experimental $C_{\rm mag}$.}
\label{Fig9}
\end{figure*}
We have also calculated the $g$-tensors from the point charge CEF calculations. For NaSrYb(BO$_3$)$_2$, we got $g_{\perp}\simeq 2.50$ and $g_{z}\simeq 2.9$, yielding an average $g$-value of $g_{\rm ave} =(2g_{\perp}+g_{z})/3 \simeq 2.63$, which is close to the value obtained from the low-temperature magnetization analysis. Similarly, the calculated $g$-components for K$_3$YbSi$_2$O$_7$ are $g_{\perp}\simeq 2.63$ and $g_{z}\simeq 2.68$. The average value $g_{\rm ave} =(2g_{\perp}+g_{z})/3 \simeq 2.65$ is also in close agreement with the one obtained from the low-temperature magnetization analysis. Further, almost equal values $g_{\perp}$ and $g_{z}$ indicates a negligibly small magnetic anisotropy in these compounds.

Using eigenvalues of the CEF Hamiltonian, we have calculated the crystal field magnetic susceptibility ($\chi_{\rm CEF}$ vs $T$) and magnetization isotherms ($M_{\rm CEF}$ vs $H$). The magnetization is estimated by calculating the expectation value of total angular momentum ($\hat{J}$) with components $J_x$, $J_y$, and $J_z$ as
% \begin{equation}
% 	M_{\rm CEF}(T,H) = \frac{N_{\rm A}g\mu_B}{Z}\sum_ke^{-\frac{E_k(H)}{k_{\rm B}T}}\left\langle\psi_k(H)|\hat{J}_{\alpha~ =~ x,~ y,~ z}|\psi_k(H)\rangle\right.
% 	\label{M_CEF}
% \end{equation}
\begin{align}
	\begin{split}
M_{\rm CEF}(T,H) = \frac{N_{\rm A}g\mu_B}{Z} ~~~~~~~~~~~~~~~~~~&\\ \times \sum_ke^{-\frac{E_k(H)}{k_{\rm B}T}}\langle\psi_k(H)|\hat{J}_{\alpha~ =~ x,~ y,~ z}|\psi_k(H)\rangle~.
	\label{M_CEF}
	\end{split}
\end{align}
Here, $Z = \sum_ke^{-E_k(H)/k_{\rm B}T}$ is the partition function, where the summation is taken over all the energy states. $\chi_{\rm CEF}$ can be calculated by taking the first derivative of $M(T,H)$ with respect to $H$. Figures~\ref{Fig8}(a) and (b) present the calculated $\chi_{\rm CEF}$ and $M_{\rm CEF}$ for NaSrYb(BO$_3$)$_2$ along with the experimental data. In both cases, we added the Van-Vleck contribution ($\chi_{\rm VV}$ and $M_{\rm VV}$) to the calculated data and our simulation reproduces the experimental data very well. Similarly, we have also calculated the $\chi_{\rm CEF}$ and $M_{\rm CEF}$ for K$_3$YbSi$_2$O$_7$, which are in very good agreement with the experimental data [see Figs.~\ref{Fig9}(a) and (b)].

The crystal field specific heat ($C_{\rm CEF}$) as a function of temperature in different fields are calculated taking into account the contributions from all the Zeeman split CEF levels. For an $N$ level system, $C_{\rm CEF}$ can be written as
%\begin{equation}
%	C_{\rm CEF}(T, H) = \frac{R}{(Zk_{\rm B}T)^2}\sum_{n>m}^{N}[E_n(H) - E_m(H)]^2e^{-\frac{[E_n(H) + E_m(H)]}{k_{\rm B}T}}.
 %\label{C_CEF}
%\end{equation}
	\begin{align}
	\begin{split}
C_{\rm CEF}(T, H) = \frac{R}{(Zk_{\rm B}T)^2} ~~~~~~~~~~~~~~~~~&\\ \times \sum_{n>m}^{N}[E_n(H) - E_m(H)]^2e^{-\frac{[E_n(H) + E_m(H)]}{k_{\rm B}T}}~.
\label{C_CEF}
\end{split}
\end{align}
Here, $E_n$ and $E_m$ are the energy of the $n^{th}$ and $m^{th}$ CEF levels, respectively. The calculated $C_{\rm CEF}$ reproduce the broad maximum in experimental $C_{\rm mag}$ of both compounds very well in the low-temperature regime [see Figs.~\ref{Fig8}(c) and~\ref{Fig9}(c)]. As we have observed in Figs.~\ref{Fig5}(b) and \ref{Fig6}(b), the broad maximum in $C_{\rm mag}(T)$ (in the presence of an applied field) can also be reproduced well by a simple two-level model, reflecting that only the Zeeman split ground state doublet contributes at low temperatures and the effect from the excited CEF levels is negligible. This is also evident from Fig.~\ref{Fig7} that the ground state of both the compounds is well separated from the first excited state ($\sim 265$~K and 370~K for NaSrYb(BO$_3$)$_2$ and K$_3$YbSi$_2$O$_7$, respectively), comprehending our assessment of $J_{\rm eff} = 1/2$ ground state.

\section{Discussion}
The low-temperature properties of NaSrYb(BO$_3$)$_2$ and K$_3$YbSi$_2$O$_7$ are described by $J_{\rm eff} = 1/2$ ground state with Kramers' doublet, which is typical for the Yb$^{3+}$ based systems~\cite{Mohanty134408}. The obtained values of $\theta_{\rm CW}^{\rm LT}$ for both the systems are very small, implying negligible magnetic interaction among the Yb$^{3+}$ ions. The energy scale of the dipolar interaction between the nearest-neighbor moments is estimated to be $D = \frac{2\mu_{0}\mu_{\rm sat}^2}{4\pi d^3} \simeq 0.022$~K for NaSrYb(BO$_3$)$_2$ and $\sim 0.014$~K for K$_3$YbSi$_2$O$_7$~\cite{Xiang224429,Guchhait144434}. Here, $d$ is the distance between NN Yb$^{3+}$ ions, $\mu_0$ is the permeability of free space. From the value of $\theta_{\rm CW}^{\rm LT}$ one can estimate the average value of nearest-neighbour exchange coupling ($J_{\rm NN}/k_{\rm B}$) as $\theta_{\rm CW}^{\rm LT} = -zJ_{\rm eff}(J_{\rm eff}+1)J_{\rm NN}/3k_{\rm B}$]~\cite{Sebastian104425}. Taking the magnetic coordination number $z = 6$ for a triangular lattice and using the experimental value of $\theta_{\rm CW}^{\rm LT}$ we got $J_{\rm NN} \simeq 0.06$~K for both the compounds. This value of $J_{\rm NN}$ is slightly larger than the dipolar interaction, the effect of which can only be observed at extremely low temperatures ($T < 0.1$~K). Interestingly, there is a site mixing of Na$^{+}$ and Sr$^{2+}$ ions that may provide randomness in the exchange interactions, an essential ingredient for suppressing magnetic LRO and holds propensity to entertain QSL~\cite{Li107202}.

For both compounds, the ground-state doublet wave function has substantial weightages coming from $\left|\pm\frac{1}{2}\right\rangle$, $\left|\pm\frac{5}{2}\right\rangle$, and $\left|\pm\frac{7}{2}\right\rangle$ states. While a large coefficient of $\left|\pm\frac{1}{2}\right\rangle$ implies significant quantum effect, the classical behaviour of the ground state is also equally probable since $\left|\pm\frac{5}{2}\right\rangle$ and $\left|\pm\frac{7}{2}\right\rangle$ also contribute to the ground state wavefunctions. Similar physics is also reported in other rare-earth based systems~\cite{Scheie356}.

Moreover, because of their frustrated geometry, weak exchange couplings, and chemical stability in 
ultra-high vacuum and high temperatures, these compounds are potential materials for ADR technique to achieve temperatures as low as few mili-Kelvins~\cite{Tokiwa42,Treu013001}. This is indeed evident from the entropy change between 0 and 9~T in Fig.~\ref{Fig5}(c).

\section{Summary}
In summary, we present magnetization, specific heat, and crystal field analysis of Yb-based triangular lattice compounds NaSrYb(BO$_3$)$_2$ and K$_3$YbSi$_2$O$_7$. There is no evidence of magnetic LRO or spin freezing down to 0.4~K for both the compounds. Thermodynamic data suggest $J_{\rm eff} = 1/2$ ground state due to Kramers' doublet, which is also reflected from point-charge CEF calculations. Due to very weak interaction among the $J_{\rm eff} = 1/2$ spins, the bulk magnetic properties are predicted by the point charge CEF model calculations only (without a magnetic term) which qualitatively reproduce our experimental results. The inelastic neutron scattering (INS) and Raman scattering experiments would be useful in order to confirm the proposed CEF scheme. Furthermore, these two materials seem to be a good alternative to hydrated paramagnetic salts for ADR applications to attain sub-Kelvin temperatures.

\section {Acknowledgments}
We would like to acknowledge SERB, India, for financial support bearing sanction Grant No.~CRG/2019/000960. SG was supported by the Prime Minister’s Research Fellowship (PMRF) scheme, Government of India.

%\bibliography{ref-NaSrYb(BO3)3}

\begin{thebibliography}{51}%
	\makeatletter
	\providecommand \@ifxundefined [1]{%
		\@ifx{#1\undefined}
	}%
	\providecommand \@ifnum [1]{%
		\ifnum #1\expandafter \@firstoftwo
		\else \expandafter \@secondoftwo
		\fi
	}%
	\providecommand \@ifx [1]{%
		\ifx #1\expandafter \@firstoftwo
		\else \expandafter \@secondoftwo
		\fi
	}%
	\providecommand \natexlab [1]{#1}%
	\providecommand \enquote  [1]{``#1''}%
	\providecommand \bibnamefont  [1]{#1}%
	\providecommand \bibfnamefont [1]{#1}%
	\providecommand \citenamefont [1]{#1}%
	\providecommand \href@noop [0]{\@secondoftwo}%
	\providecommand \href [0]{\begingroup \@sanitize@url \@href}%
	\providecommand \@href[1]{\@@startlink{#1}\@@href}%
	\providecommand \@@href[1]{\endgroup#1\@@endlink}%
	\providecommand \@sanitize@url [0]{\catcode `\\12\catcode `\$12\catcode
		`\&12\catcode `\#12\catcode `\^12\catcode `\_12\catcode `\%12\relax}%
	\providecommand \@@startlink[1]{}%
	\providecommand \@@endlink[0]{}%
	\providecommand \url  [0]{\begingroup\@sanitize@url \@url }%
	\providecommand \@url [1]{\endgroup\@href {#1}{\urlprefix }}%
	\providecommand \urlprefix  [0]{URL }%
	\providecommand \Eprint [0]{\href }%
	\providecommand \doibase [0]{https://doi.org/}%
	\providecommand \selectlanguage [0]{\@gobble}%
	\providecommand \bibinfo  [0]{\@secondoftwo}%
	\providecommand \bibfield  [0]{\@secondoftwo}%
	\providecommand \translation [1]{[#1]}%
	\providecommand \BibitemOpen [0]{}%
	\providecommand \bibitemStop [0]{}%
	\providecommand \bibitemNoStop [0]{.\EOS\space}%
	\providecommand \EOS [0]{\spacefactor3000\relax}%
	\providecommand \BibitemShut  [1]{\csname bibitem#1\endcsname}%
	\let\auto@bib@innerbib\@empty
	%</preamble>
	\bibitem [{\citenamefont {Savary}\ and\ \citenamefont
		{Balents}(2016)}]{Savary016502}%
	\BibitemOpen
	\bibfield  {author} {\bibinfo {author} {\bibfnamefont {L.}~\bibnamefont
			{Savary}}\ and\ \bibinfo {author} {\bibfnamefont {L.}~\bibnamefont
			{Balents}},\ }\bibfield  {title} {\bibinfo {title} {Quantum spin liquids: a
			review},\ }\href {https://doi.org/10.1088/0034-4885/80/1/016502} {\bibfield
		{journal} {\bibinfo  {journal} {Rep. Prog. Phys.}\ }\textbf {\bibinfo
			{volume} {80}},\ \bibinfo {pages} {016502} (\bibinfo {year}
		{2016})}\BibitemShut {NoStop}%
	\bibitem [{\citenamefont {Balents}(2010)}]{Balents199}%
	\BibitemOpen
	\bibfield  {author} {\bibinfo {author} {\bibfnamefont {L.}~\bibnamefont
			{Balents}},\ }\bibfield  {title} {\bibinfo {title} {Spin liquids in
			frustrated magnets},\ }\href {https://doi.org/10.1038/nature08917} {\bibfield
		{journal} {\bibinfo  {journal} {Nature}\ }\textbf {\bibinfo {volume}
			{464}},\ \bibinfo {pages} {199} (\bibinfo {year} {2010})}\BibitemShut
	{NoStop}%
	\bibitem [{\citenamefont {Anderson}(1973)}]{Anderson153}%
	\BibitemOpen
	\bibfield  {author} {\bibinfo {author} {\bibfnamefont {P.}~\bibnamefont
			{Anderson}},\ }\bibfield  {title} {\bibinfo {title} {Resonating valence
			bonds: A new kind of insulator?},\ }\href
	{https://doi.org/https://doi.org/10.1016/0025-5408(73)90167-0} {\bibfield
		{journal} {\bibinfo  {journal} {Mater. Res. Bull.}\ }\textbf {\bibinfo
			{volume} {8}},\ \bibinfo {pages} {153} (\bibinfo {year} {1973})}\BibitemShut
	{NoStop}%
	\bibitem [{\citenamefont {Singh}\ and\ \citenamefont {Huse}(1992)}]{Singh1766}%
	\BibitemOpen
	\bibfield  {author} {\bibinfo {author} {\bibfnamefont {R.~R.~P.}\
			\bibnamefont {Singh}}\ and\ \bibinfo {author} {\bibfnamefont {D.~A.}\
			\bibnamefont {Huse}},\ }\bibfield  {title} {\bibinfo {title}
		{{Three-sublattice order in triangular- and Kagom\'e-lattice spin-half
				antiferromagnets}},\ }\href {https://doi.org/10.1103/PhysRevLett.68.1766}
	{\bibfield  {journal} {\bibinfo  {journal} {Phys. Rev. Lett.}\ }\textbf
		{\bibinfo {volume} {68}},\ \bibinfo {pages} {1766} (\bibinfo {year}
		{1992})}\BibitemShut {NoStop}%
	\bibitem [{\citenamefont {Capriotti}\ \emph {et~al.}(1999)\citenamefont
		{Capriotti}, \citenamefont {Trumper},\ and\ \citenamefont
		{Sorella}}]{Capriotti3899}%
	\BibitemOpen
	\bibfield  {author} {\bibinfo {author} {\bibfnamefont {L.}~\bibnamefont
			{Capriotti}}, \bibinfo {author} {\bibfnamefont {A.~E.}\ \bibnamefont
			{Trumper}},\ and\ \bibinfo {author} {\bibfnamefont {S.}~\bibnamefont
			{Sorella}},\ }\bibfield  {title} {\bibinfo {title} {{Long-Range N\'eel Order
				in the Triangular Heisenberg Model}},\ }\href
	{https://doi.org/10.1103/PhysRevLett.82.3899} {\bibfield  {journal} {\bibinfo
			{journal} {Phys. Rev. Lett.}\ }\textbf {\bibinfo {volume} {82}},\ \bibinfo
		{pages} {3899} (\bibinfo {year} {1999})}\BibitemShut {NoStop}%
	\bibitem [{\citenamefont {Zhu}\ \emph {et~al.}(2018)\citenamefont {Zhu},
		\citenamefont {Maksimov}, \citenamefont {White},\ and\ \citenamefont
		{Chernyshev}}]{Zhu207203}%
	\BibitemOpen
	\bibfield  {author} {\bibinfo {author} {\bibfnamefont {Z.}~\bibnamefont
			{Zhu}}, \bibinfo {author} {\bibfnamefont {P.~A.}\ \bibnamefont {Maksimov}},
		\bibinfo {author} {\bibfnamefont {S.~R.}\ \bibnamefont {White}},\ and\
		\bibinfo {author} {\bibfnamefont {A.~L.}\ \bibnamefont {Chernyshev}},\
	}\bibfield  {title} {\bibinfo {title} {{Topography of Spin Liquids on a
				Triangular Lattice}},\ }\href
	{https://doi.org/10.1103/PhysRevLett.120.207203} {\bibfield  {journal}
		{\bibinfo  {journal} {Phys. Rev. Lett.}\ }\textbf {\bibinfo {volume} {120}},\
		\bibinfo {pages} {207203} (\bibinfo {year} {2018})}\BibitemShut {NoStop}%
	\bibitem [{\citenamefont {Drescher}\ \emph {et~al.}(2023)\citenamefont
		{Drescher}, \citenamefont {Vanderstraeten}, \citenamefont {Moessner},\ and\
		\citenamefont {Pollmann}}]{Drescher220401}%
	\BibitemOpen
	\bibfield  {author} {\bibinfo {author} {\bibfnamefont {M.}~\bibnamefont
			{Drescher}}, \bibinfo {author} {\bibfnamefont {L.}~\bibnamefont
			{Vanderstraeten}}, \bibinfo {author} {\bibfnamefont {R.}~\bibnamefont
			{Moessner}},\ and\ \bibinfo {author} {\bibfnamefont {F.}~\bibnamefont
			{Pollmann}},\ }\bibfield  {title} {\bibinfo {title} {{Dynamical signatures of
				symmetry-broken and liquid phases in an $S$ = $\frac{1}{2}$ Heisenberg
				antiferromagnet on the triangular lattice}},\ }\href
	{https://doi.org/10.1103/PhysRevB.108.L220401} {\bibfield  {journal}
		{\bibinfo  {journal} {Phys. Rev. B}\ }\textbf {\bibinfo {volume} {108}},\
		\bibinfo {pages} {L220401} (\bibinfo {year} {2023})}\BibitemShut {NoStop}%
	\bibitem [{\citenamefont {Zhu}\ and\ \citenamefont {White}(2015)}]{Zhu041105}%
	\BibitemOpen
	\bibfield  {author} {\bibinfo {author} {\bibfnamefont {Z.}~\bibnamefont
			{Zhu}}\ and\ \bibinfo {author} {\bibfnamefont {S.~R.}\ \bibnamefont
			{White}},\ }\bibfield  {title} {\bibinfo {title} {{Spin liquid phase of the
				$S=\frac{1}{2}\phantom{\rule{4.pt}{0ex}}{J}_{1}\ensuremath{-}{J}_{2}$
				Heisenberg model on the triangular lattice}},\ }\href
	{https://doi.org/10.1103/PhysRevB.92.041105} {\bibfield  {journal} {\bibinfo
			{journal} {Phys. Rev. B}\ }\textbf {\bibinfo {volume} {92}},\ \bibinfo
		{pages} {041105} (\bibinfo {year} {2015})}\BibitemShut {NoStop}%
	\bibitem [{\citenamefont {Hu}\ \emph {et~al.}(2015)\citenamefont {Hu},
		\citenamefont {Gong}, \citenamefont {Zhu},\ and\ \citenamefont
		{Sheng}}]{Hu140403}%
	\BibitemOpen
	\bibfield  {author} {\bibinfo {author} {\bibfnamefont {W.-J.}\ \bibnamefont
			{Hu}}, \bibinfo {author} {\bibfnamefont {S.-S.}\ \bibnamefont {Gong}},
		\bibinfo {author} {\bibfnamefont {W.}~\bibnamefont {Zhu}},\ and\ \bibinfo
		{author} {\bibfnamefont {D.~N.}\ \bibnamefont {Sheng}},\ }\bibfield  {title}
	{\bibinfo {title} {{Competing spin-liquid states in the spin-$\frac{1}{2}$
				Heisenberg model on the triangular lattice}},\ }\href
	{https://doi.org/10.1103/PhysRevB.92.140403} {\bibfield  {journal} {\bibinfo
			{journal} {Phys. Rev. B}\ }\textbf {\bibinfo {volume} {92}},\ \bibinfo
		{pages} {140403} (\bibinfo {year} {2015})}\BibitemShut {NoStop}%
	\bibitem [{\citenamefont {Kimchi}\ \emph {et~al.}(2018)\citenamefont {Kimchi},
		\citenamefont {Nahum},\ and\ \citenamefont {Senthil}}]{Kimchi031028}%
	\BibitemOpen
	\bibfield  {author} {\bibinfo {author} {\bibfnamefont {I.}~\bibnamefont
			{Kimchi}}, \bibinfo {author} {\bibfnamefont {A.}~\bibnamefont {Nahum}},\ and\
		\bibinfo {author} {\bibfnamefont {T.}~\bibnamefont {Senthil}},\ }\bibfield
	{title} {\bibinfo {title} {{Valence Bonds in Random Quantum Magnets: Theory
				and Application to ${\mathrm{YbMgGaO}}_{4}$}},\ }\href
	{https://doi.org/10.1103/PhysRevX.8.031028} {\bibfield  {journal} {\bibinfo
			{journal} {Phys. Rev. X}\ }\textbf {\bibinfo {volume} {8}},\ \bibinfo {pages}
		{031028} (\bibinfo {year} {2018})}\BibitemShut {NoStop}%
	\bibitem [{\citenamefont {Li}\ \emph {et~al.}(2017{\natexlab{a}})\citenamefont
		{Li}, \citenamefont {Adroja}, \citenamefont {Bewley}, \citenamefont
		{Voneshen}, \citenamefont {Tsirlin}, \citenamefont {Gegenwart},\ and\
		\citenamefont {Zhang}}]{Li107202}%
	\BibitemOpen
	\bibfield  {author} {\bibinfo {author} {\bibfnamefont {Y.}~\bibnamefont
			{Li}}, \bibinfo {author} {\bibfnamefont {D.}~\bibnamefont {Adroja}}, \bibinfo
		{author} {\bibfnamefont {R.~I.}\ \bibnamefont {Bewley}}, \bibinfo {author}
		{\bibfnamefont {D.}~\bibnamefont {Voneshen}}, \bibinfo {author}
		{\bibfnamefont {A.~A.}\ \bibnamefont {Tsirlin}}, \bibinfo {author}
		{\bibfnamefont {P.}~\bibnamefont {Gegenwart}},\ and\ \bibinfo {author}
		{\bibfnamefont {Q.}~\bibnamefont {Zhang}},\ }\bibfield  {title} {\bibinfo
		{title} {{Crystalline Electric-Field Randomness in the Triangular Lattice
				Spin-Liquid ${\mathrm{YbMgGaO}}_{4}$}},\ }\href
	{https://doi.org/10.1103/PhysRevLett.118.107202} {\bibfield  {journal}
		{\bibinfo  {journal} {Phys. Rev. Lett.}\ }\textbf {\bibinfo {volume} {118}},\
		\bibinfo {pages} {107202} (\bibinfo {year} {2017}{\natexlab{a}})}\BibitemShut
	{NoStop}%
	\bibitem [{\citenamefont {Kundu}\ \emph {et~al.}(2020)\citenamefont {Kundu},
		\citenamefont {Hossain}, \citenamefont {S.}, \citenamefont {Das},
		\citenamefont {Baenitz}, \citenamefont {Baker}, \citenamefont {Orain},
		\citenamefont {Joshi}, \citenamefont {Mathieu}, \citenamefont {Mahadevan},
		\citenamefont {Pujari}, \citenamefont {Bhattacharjee}, \citenamefont
		{Mahajan},\ and\ \citenamefont {Sarma}}]{Kundu117206}%
	\BibitemOpen
	\bibfield  {author} {\bibinfo {author} {\bibfnamefont {S.}~\bibnamefont
			{Kundu}}, \bibinfo {author} {\bibfnamefont {A.}~\bibnamefont {Hossain}},
		\bibinfo {author} {\bibfnamefont {P.~K.}\ \bibnamefont {S.}}, \bibinfo
		{author} {\bibfnamefont {R.}~\bibnamefont {Das}}, \bibinfo {author}
		{\bibfnamefont {M.}~\bibnamefont {Baenitz}}, \bibinfo {author} {\bibfnamefont
			{P.~J.}\ \bibnamefont {Baker}}, \bibinfo {author} {\bibfnamefont {J.-C.}\
			\bibnamefont {Orain}}, \bibinfo {author} {\bibfnamefont {D.~C.}\ \bibnamefont
			{Joshi}}, \bibinfo {author} {\bibfnamefont {R.}~\bibnamefont {Mathieu}},
		\bibinfo {author} {\bibfnamefont {P.}~\bibnamefont {Mahadevan}}, \bibinfo
		{author} {\bibfnamefont {S.}~\bibnamefont {Pujari}}, \bibinfo {author}
		{\bibfnamefont {S.}~\bibnamefont {Bhattacharjee}}, \bibinfo {author}
		{\bibfnamefont {A.~V.}\ \bibnamefont {Mahajan}},\ and\ \bibinfo {author}
		{\bibfnamefont {D.~D.}\ \bibnamefont {Sarma}},\ }\bibfield  {title} {\bibinfo
		{title} {{Signatures of a Spin-$\frac{1}{2}$ Cooperative Paramagnet in the
				Diluted Triangular Lattice of ${\mathrm{Y}}_{2}{\mathrm{CuTiO}}_{6}$}},\
	}\href {https://doi.org/10.1103/PhysRevLett.125.117206} {\bibfield  {journal}
		{\bibinfo  {journal} {Phys. Rev. Lett.}\ }\textbf {\bibinfo {volume} {125}},\
		\bibinfo {pages} {117206} (\bibinfo {year} {2020})}\BibitemShut {NoStop}%
	\bibitem [{\citenamefont {Zhu}\ \emph {et~al.}(2017)\citenamefont {Zhu},
		\citenamefont {Maksimov}, \citenamefont {White},\ and\ \citenamefont
		{Chernyshev}}]{Zhu157201}%
	\BibitemOpen
	\bibfield  {author} {\bibinfo {author} {\bibfnamefont {Z.}~\bibnamefont
			{Zhu}}, \bibinfo {author} {\bibfnamefont {P.~A.}\ \bibnamefont {Maksimov}},
		\bibinfo {author} {\bibfnamefont {S.~R.}\ \bibnamefont {White}},\ and\
		\bibinfo {author} {\bibfnamefont {A.~L.}\ \bibnamefont {Chernyshev}},\
	}\bibfield  {title} {\bibinfo {title} {{Disorder-Induced Mimicry of a Spin
				Liquid in ${\mathrm{YbMgGaO}}_{4}$}},\ }\href
	{https://doi.org/10.1103/PhysRevLett.119.157201} {\bibfield  {journal}
		{\bibinfo  {journal} {Phys. Rev. Lett.}\ }\textbf {\bibinfo {volume} {119}},\
		\bibinfo {pages} {157201} (\bibinfo {year} {2017})}\BibitemShut {NoStop}%
	\bibitem [{\citenamefont {Somesh}\ \emph {et~al.}(2021)\citenamefont {Somesh},
		\citenamefont {Furukawa}, \citenamefont {Simutis}, \citenamefont {Bert},
		\citenamefont {Prinz-Zwick}, \citenamefont {B\"uttgen}, \citenamefont
		{Zorko}, \citenamefont {Tsirlin}, \citenamefont {Mendels},\ and\
		\citenamefont {Nath}}]{Somesh104422}%
	\BibitemOpen
	\bibfield  {author} {\bibinfo {author} {\bibfnamefont {K.}~\bibnamefont
			{Somesh}}, \bibinfo {author} {\bibfnamefont {Y.}~\bibnamefont {Furukawa}},
		\bibinfo {author} {\bibfnamefont {G.}~\bibnamefont {Simutis}}, \bibinfo
		{author} {\bibfnamefont {F.}~\bibnamefont {Bert}}, \bibinfo {author}
		{\bibfnamefont {M.}~\bibnamefont {Prinz-Zwick}}, \bibinfo {author}
		{\bibfnamefont {N.}~\bibnamefont {B\"uttgen}}, \bibinfo {author}
		{\bibfnamefont {A.}~\bibnamefont {Zorko}}, \bibinfo {author} {\bibfnamefont
			{A.~A.}\ \bibnamefont {Tsirlin}}, \bibinfo {author} {\bibfnamefont
			{P.}~\bibnamefont {Mendels}},\ and\ \bibinfo {author} {\bibfnamefont
			{R.}~\bibnamefont {Nath}},\ }\bibfield  {title} {\bibinfo {title} {Universal
			fluctuating regime in triangular chromate antiferromagnets},\ }\href
	{https://doi.org/10.1103/PhysRevB.104.104422} {\bibfield  {journal} {\bibinfo
			{journal} {Phys. Rev. B}\ }\textbf {\bibinfo {volume} {104}},\ \bibinfo
		{pages} {104422} (\bibinfo {year} {2021})}\BibitemShut {NoStop}%
	\bibitem [{\citenamefont {Lal}\ \emph {et~al.}(2023)\citenamefont {Lal},
		\citenamefont {Sebastian}, \citenamefont {Islam}, \citenamefont {Saravanan},
		\citenamefont {Uhlarz}, \citenamefont {Skourski},\ and\ \citenamefont
		{Nath}}]{Lal014429}%
	\BibitemOpen
	\bibfield  {author} {\bibinfo {author} {\bibfnamefont {S.}~\bibnamefont
			{Lal}}, \bibinfo {author} {\bibfnamefont {S.~J.}\ \bibnamefont {Sebastian}},
		\bibinfo {author} {\bibfnamefont {S.~S.}\ \bibnamefont {Islam}}, \bibinfo
		{author} {\bibfnamefont {M.~P.}\ \bibnamefont {Saravanan}}, \bibinfo {author}
		{\bibfnamefont {M.}~\bibnamefont {Uhlarz}}, \bibinfo {author} {\bibfnamefont
			{Y.}~\bibnamefont {Skourski}},\ and\ \bibinfo {author} {\bibfnamefont
			{R.}~\bibnamefont {Nath}},\ }\bibfield  {title} {\bibinfo {title} {{{Double
					magnetic transitions and exotic field-induced phase in the triangular lattice
					antiferromagnets
					${\mathrm{Sr}}_{3}\mathrm{Co}{(\mathrm{Nb},\mathrm{Ta})}_{2}{\mathrm{O}}_{9}$}}},\
	}\href {https://doi.org/10.1103/PhysRevB.108.014429} {\bibfield  {journal}
		{\bibinfo  {journal} {Phys. Rev. B}\ }\textbf {\bibinfo {volume} {108}},\
		\bibinfo {pages} {014429} (\bibinfo {year} {2023})}\BibitemShut {NoStop}%
	\bibitem [{\citenamefont {Li}\ \emph {et~al.}(2016)\citenamefont {Li},
		\citenamefont {Adroja}, \citenamefont {Biswas}, \citenamefont {Baker},
		\citenamefont {Zhang}, \citenamefont {Liu}, \citenamefont {Tsirlin},
		\citenamefont {Gegenwart},\ and\ \citenamefont {Zhang}}]{Li097201}%
	\BibitemOpen
	\bibfield  {author} {\bibinfo {author} {\bibfnamefont {Y.}~\bibnamefont
			{Li}}, \bibinfo {author} {\bibfnamefont {D.}~\bibnamefont {Adroja}}, \bibinfo
		{author} {\bibfnamefont {P.~K.}\ \bibnamefont {Biswas}}, \bibinfo {author}
		{\bibfnamefont {P.~J.}\ \bibnamefont {Baker}}, \bibinfo {author}
		{\bibfnamefont {Q.}~\bibnamefont {Zhang}}, \bibinfo {author} {\bibfnamefont
			{J.}~\bibnamefont {Liu}}, \bibinfo {author} {\bibfnamefont {A.~A.}\
			\bibnamefont {Tsirlin}}, \bibinfo {author} {\bibfnamefont {P.}~\bibnamefont
			{Gegenwart}},\ and\ \bibinfo {author} {\bibfnamefont {Q.}~\bibnamefont
			{Zhang}},\ }\bibfield  {title} {\bibinfo {title} {{Muon Spin Relaxation
				Evidence for the U(1) Quantum Spin-Liquid Ground State in the Triangular
				Antiferromagnet YbMgGaO$_4$}},\ }\href
	{https://doi.org/10.1103/PhysRevLett.117.097201} {\bibfield  {journal}
		{\bibinfo  {journal} {Phys. Rev. Lett.}\ }\textbf {\bibinfo {volume} {117}},\
		\bibinfo {pages} {097201} (\bibinfo {year} {2016})}\BibitemShut {NoStop}%
	\bibitem [{\citenamefont {Dai}\ \emph {et~al.}(2021)\citenamefont {Dai},
		\citenamefont {Zhang}, \citenamefont {Xie}, \citenamefont {Duan},
		\citenamefont {Gao}, \citenamefont {Zhu}, \citenamefont {Feng}, \citenamefont
		{Tao}, \citenamefont {Huang}, \citenamefont {Cao}, \citenamefont
		{Podlesnyak}, \citenamefont {Granroth}, \citenamefont {Everett},
		\citenamefont {Neuefeind}, \citenamefont {Voneshen}, \citenamefont {Wang},
		\citenamefont {Tan}, \citenamefont {Morosan}, \citenamefont {Wang},
		\citenamefont {Lin}, \citenamefont {Shu}, \citenamefont {Chen}, \citenamefont
		{Guo}, \citenamefont {Lu},\ and\ \citenamefont {Dai}}]{Dai021044}%
	\BibitemOpen
	\bibfield  {author} {\bibinfo {author} {\bibfnamefont {P.-L.}\ \bibnamefont
			{Dai}}, \bibinfo {author} {\bibfnamefont {G.}~\bibnamefont {Zhang}}, \bibinfo
		{author} {\bibfnamefont {Y.}~\bibnamefont {Xie}}, \bibinfo {author}
		{\bibfnamefont {C.}~\bibnamefont {Duan}}, \bibinfo {author} {\bibfnamefont
			{Y.}~\bibnamefont {Gao}}, \bibinfo {author} {\bibfnamefont {Z.}~\bibnamefont
			{Zhu}}, \bibinfo {author} {\bibfnamefont {E.}~\bibnamefont {Feng}}, \bibinfo
		{author} {\bibfnamefont {Z.}~\bibnamefont {Tao}}, \bibinfo {author}
		{\bibfnamefont {C.-L.}\ \bibnamefont {Huang}}, \bibinfo {author}
		{\bibfnamefont {H.}~\bibnamefont {Cao}}, \bibinfo {author} {\bibfnamefont
			{A.}~\bibnamefont {Podlesnyak}}, \bibinfo {author} {\bibfnamefont {G.~E.}\
			\bibnamefont {Granroth}}, \bibinfo {author} {\bibfnamefont {M.~S.}\
			\bibnamefont {Everett}}, \bibinfo {author} {\bibfnamefont {J.~C.}\
			\bibnamefont {Neuefeind}}, \bibinfo {author} {\bibfnamefont {D.}~\bibnamefont
			{Voneshen}}, \bibinfo {author} {\bibfnamefont {S.}~\bibnamefont {Wang}},
		\bibinfo {author} {\bibfnamefont {G.}~\bibnamefont {Tan}}, \bibinfo {author}
		{\bibfnamefont {E.}~\bibnamefont {Morosan}}, \bibinfo {author} {\bibfnamefont
			{X.}~\bibnamefont {Wang}}, \bibinfo {author} {\bibfnamefont {H.-Q.}\
			\bibnamefont {Lin}}, \bibinfo {author} {\bibfnamefont {L.}~\bibnamefont
			{Shu}}, \bibinfo {author} {\bibfnamefont {G.}~\bibnamefont {Chen}}, \bibinfo
		{author} {\bibfnamefont {Y.}~\bibnamefont {Guo}}, \bibinfo {author}
		{\bibfnamefont {X.}~\bibnamefont {Lu}},\ and\ \bibinfo {author}
		{\bibfnamefont {P.}~\bibnamefont {Dai}},\ }\bibfield  {title} {\bibinfo
		{title} {{Spinon Fermi Surface Spin Liquid in a Triangular Lattice
				Antiferromagnet ${\mathrm{NaYbSe}}_{2}$}},\ }\href
	{https://doi.org/10.1103/PhysRevX.11.021044} {\bibfield  {journal} {\bibinfo
			{journal} {Phys. Rev. X}\ }\textbf {\bibinfo {volume} {11}},\ \bibinfo
		{pages} {021044} (\bibinfo {year} {2021})}\BibitemShut {NoStop}%
	\bibitem [{\citenamefont {Li}\ \emph {et~al.}(2017{\natexlab{b}})\citenamefont
		{Li}, \citenamefont {Adroja}, \citenamefont {Voneshen}, \citenamefont
		{Bewley}, \citenamefont {Zhang}, \citenamefont {Tsirlin},\ and\ \citenamefont
		{Gegenwart}}]{Li15814}%
	\BibitemOpen
	\bibfield  {author} {\bibinfo {author} {\bibfnamefont {Y.}~\bibnamefont
			{Li}}, \bibinfo {author} {\bibfnamefont {D.}~\bibnamefont {Adroja}}, \bibinfo
		{author} {\bibfnamefont {D.}~\bibnamefont {Voneshen}}, \bibinfo {author}
		{\bibfnamefont {R.~I.}\ \bibnamefont {Bewley}}, \bibinfo {author}
		{\bibfnamefont {Q.}~\bibnamefont {Zhang}}, \bibinfo {author} {\bibfnamefont
			{A.~A.}\ \bibnamefont {Tsirlin}},\ and\ \bibinfo {author} {\bibfnamefont
			{P.}~\bibnamefont {Gegenwart}},\ }\bibfield  {title} {\bibinfo {title}
		{{Nearest-neighbour resonating valence bonds in YbMgGaO$_4$}},\ }\href
	{https://doi.org/10.1038/ncomms15814} {\bibfield  {journal} {\bibinfo
			{journal} {Nat. Commun.}\ }\textbf {\bibinfo {volume} {8}},\ \bibinfo {pages}
		{15814} (\bibinfo {year} {2017}{\natexlab{b}})}\BibitemShut {NoStop}%
	\bibitem [{\citenamefont {Sarkar}\ \emph {et~al.}(2019)\citenamefont {Sarkar},
		\citenamefont {Schlender}, \citenamefont {Grinenko}, \citenamefont
		{Haeussler}, \citenamefont {Baker}, \citenamefont {Doert},\ and\
		\citenamefont {Klauss}}]{Sarkar241116}%
	\BibitemOpen
	\bibfield  {author} {\bibinfo {author} {\bibfnamefont {R.}~\bibnamefont
			{Sarkar}}, \bibinfo {author} {\bibfnamefont {P.}~\bibnamefont {Schlender}},
		\bibinfo {author} {\bibfnamefont {V.}~\bibnamefont {Grinenko}}, \bibinfo
		{author} {\bibfnamefont {E.}~\bibnamefont {Haeussler}}, \bibinfo {author}
		{\bibfnamefont {P.~J.}\ \bibnamefont {Baker}}, \bibinfo {author}
		{\bibfnamefont {T.}~\bibnamefont {Doert}},\ and\ \bibinfo {author}
		{\bibfnamefont {H.-H.}\ \bibnamefont {Klauss}},\ }\bibfield  {title}
	{\bibinfo {title} {{Quantum spin liquid ground state in the disorder free
				triangular lattice ${\mathrm{NaYbS}}_{2}$}},\ }\href
	{https://doi.org/10.1103/PhysRevB.100.241116} {\bibfield  {journal} {\bibinfo
			{journal} {Phys. Rev. B}\ }\textbf {\bibinfo {volume} {100}},\ \bibinfo
		{pages} {241116} (\bibinfo {year} {2019})}\BibitemShut {NoStop}%
	\bibitem [{\citenamefont {Scheie}\ \emph {et~al.}(2024)\citenamefont {Scheie},
		\citenamefont {Kamiya}, \citenamefont {Zhang}, \citenamefont {Lee},
		\citenamefont {Woods}, \citenamefont {Ajeesh}, \citenamefont {Gonzalez},
		\citenamefont {Bernu}, \citenamefont {Villanova}, \citenamefont {Xing},
		\citenamefont {Huang}, \citenamefont {Zhang}, \citenamefont {Ma},
		\citenamefont {Choi}, \citenamefont {Pajerowski}, \citenamefont {Zhou},
		\citenamefont {Sefat}, \citenamefont {Okamoto}, \citenamefont {Berlijn},
		\citenamefont {Messio}, \citenamefont {Movshovich}, \citenamefont {Batista},\
		and\ \citenamefont {Tennant}}]{Scheie014425}%
	\BibitemOpen
	\bibfield  {author} {\bibinfo {author} {\bibfnamefont {A.~O.}\ \bibnamefont
			{Scheie}}, \bibinfo {author} {\bibfnamefont {Y.}~\bibnamefont {Kamiya}},
		\bibinfo {author} {\bibfnamefont {H.}~\bibnamefont {Zhang}}, \bibinfo
		{author} {\bibfnamefont {S.}~\bibnamefont {Lee}}, \bibinfo {author}
		{\bibfnamefont {A.~J.}\ \bibnamefont {Woods}}, \bibinfo {author}
		{\bibfnamefont {M.~O.}\ \bibnamefont {Ajeesh}}, \bibinfo {author}
		{\bibfnamefont {M.~G.}\ \bibnamefont {Gonzalez}}, \bibinfo {author}
		{\bibfnamefont {B.}~\bibnamefont {Bernu}}, \bibinfo {author} {\bibfnamefont
			{J.~W.}\ \bibnamefont {Villanova}}, \bibinfo {author} {\bibfnamefont
			{J.}~\bibnamefont {Xing}}, \bibinfo {author} {\bibfnamefont {Q.}~\bibnamefont
			{Huang}}, \bibinfo {author} {\bibfnamefont {Q.}~\bibnamefont {Zhang}},
		\bibinfo {author} {\bibfnamefont {J.}~\bibnamefont {Ma}}, \bibinfo {author}
		{\bibfnamefont {E.~S.}\ \bibnamefont {Choi}}, \bibinfo {author}
		{\bibfnamefont {D.~M.}\ \bibnamefont {Pajerowski}}, \bibinfo {author}
		{\bibfnamefont {H.}~\bibnamefont {Zhou}}, \bibinfo {author} {\bibfnamefont
			{A.~S.}\ \bibnamefont {Sefat}}, \bibinfo {author} {\bibfnamefont
			{S.}~\bibnamefont {Okamoto}}, \bibinfo {author} {\bibfnamefont
			{T.}~\bibnamefont {Berlijn}}, \bibinfo {author} {\bibfnamefont
			{L.}~\bibnamefont {Messio}}, \bibinfo {author} {\bibfnamefont
			{R.}~\bibnamefont {Movshovich}}, \bibinfo {author} {\bibfnamefont {C.~D.}\
			\bibnamefont {Batista}},\ and\ \bibinfo {author} {\bibfnamefont {D.~A.}\
			\bibnamefont {Tennant}},\ }\bibfield  {title} {\bibinfo {title} {{Nonlinear
				magnons and exchange Hamiltonians of the delafossite proximate quantum spin
				liquid candidates ${\text{KYbSe}}_{2}$ and ${\text{NaYbSe}}_{2}$}},\ }\href
	{https://doi.org/10.1103/PhysRevB.109.014425} {\bibfield  {journal} {\bibinfo
			{journal} {Phys. Rev. B}\ }\textbf {\bibinfo {volume} {109}},\ \bibinfo
		{pages} {014425} (\bibinfo {year} {2024})}\BibitemShut {NoStop}%
	\bibitem [{\citenamefont {Treu}\ \emph {et~al.}(2024)\citenamefont {Treu},
		\citenamefont {Klinger}, \citenamefont {Oefele}, \citenamefont {Telang},
		\citenamefont {Jesche},\ and\ \citenamefont {Gegenwart}}]{Treu013001}%
	\BibitemOpen
	\bibfield  {author} {\bibinfo {author} {\bibfnamefont {T.}~\bibnamefont
			{Treu}}, \bibinfo {author} {\bibfnamefont {M.}~\bibnamefont {Klinger}},
		\bibinfo {author} {\bibfnamefont {N.}~\bibnamefont {Oefele}}, \bibinfo
		{author} {\bibfnamefont {P.}~\bibnamefont {Telang}}, \bibinfo {author}
		{\bibfnamefont {A.}~\bibnamefont {Jesche}},\ and\ \bibinfo {author}
		{\bibfnamefont {P.}~\bibnamefont {Gegenwart}},\ }\bibfield  {title} {\bibinfo
		{title} {{Utilizing frustration in Gd- and Yb-based oxides for milli-Kelvin
				adiabatic demagnetization refrigeration}},\ }\href
	{https://doi.org/10.1088/1361-648X/ad7dc5} {\bibfield  {journal} {\bibinfo
			{journal} {J. Phys.: Condens. Matter}\ }\textbf {\bibinfo {volume} {37}},\
		\bibinfo {pages} {013001} (\bibinfo {year} {2024})}\BibitemShut {NoStop}%
	\bibitem [{\citenamefont {Tokiwa}\ \emph {et~al.}(2021)\citenamefont {Tokiwa},
		\citenamefont {Bachus}, \citenamefont {Kavita}, \citenamefont {Jesche},
		\citenamefont {Tsirlin},\ and\ \citenamefont {Gegenwart}}]{Tokiwa42}%
	\BibitemOpen
	\bibfield  {author} {\bibinfo {author} {\bibfnamefont {Y.}~\bibnamefont
			{Tokiwa}}, \bibinfo {author} {\bibfnamefont {S.}~\bibnamefont {Bachus}},
		\bibinfo {author} {\bibfnamefont {K.}~\bibnamefont {Kavita}}, \bibinfo
		{author} {\bibfnamefont {A.}~\bibnamefont {Jesche}}, \bibinfo {author}
		{\bibfnamefont {A.~A.}\ \bibnamefont {Tsirlin}},\ and\ \bibinfo {author}
		{\bibfnamefont {P.}~\bibnamefont {Gegenwart}},\ }\bibfield  {title} {\bibinfo
		{title} {Frustrated magnet for adiabatic demagnetization cooling to
			milli-kelvin temperatures},\ }\href
	{https://doi.org/10.1038/s43246-021-00142-1} {\bibfield  {journal} {\bibinfo
			{journal} {Commun. Mater.}\ }\textbf {\bibinfo {volume} {2}},\ \bibinfo
		{pages} {42} (\bibinfo {year} {2021})}\BibitemShut {NoStop}%
	\bibitem [{\citenamefont {Jesche}\ \emph {et~al.}(2023)\citenamefont {Jesche},
		\citenamefont {Winterhalter-Stocker}, \citenamefont {Hirschberger},
		\citenamefont {Bellon}, \citenamefont {Bachus}, \citenamefont {Tokiwa},
		\citenamefont {Tsirlin},\ and\ \citenamefont {Gegenwart}}]{Jesche104402}%
	\BibitemOpen
	\bibfield  {author} {\bibinfo {author} {\bibfnamefont {A.}~\bibnamefont
			{Jesche}}, \bibinfo {author} {\bibfnamefont {N.}~\bibnamefont
			{Winterhalter-Stocker}}, \bibinfo {author} {\bibfnamefont {F.}~\bibnamefont
			{Hirschberger}}, \bibinfo {author} {\bibfnamefont {A.}~\bibnamefont
			{Bellon}}, \bibinfo {author} {\bibfnamefont {S.}~\bibnamefont {Bachus}},
		\bibinfo {author} {\bibfnamefont {Y.}~\bibnamefont {Tokiwa}}, \bibinfo
		{author} {\bibfnamefont {A.~A.}\ \bibnamefont {Tsirlin}},\ and\ \bibinfo
		{author} {\bibfnamefont {P.}~\bibnamefont {Gegenwart}},\ }\bibfield  {title}
	{\bibinfo {title} {{Adiabatic demagnetization cooling well below the magnetic
				ordering temperature in the triangular antiferromagnet KBaGd(BO$_3)_2$}},\
	}\href {https://doi.org/10.1103/PhysRevB.107.104402} {\bibfield  {journal}
		{\bibinfo  {journal} {Phys. Rev. B}\ }\textbf {\bibinfo {volume} {107}},\
		\bibinfo {pages} {104402} (\bibinfo {year} {2023})}\BibitemShut {NoStop}%
	\bibitem [{\citenamefont {Xiang}\ \emph {et~al.}()\citenamefont {Xiang},
		\citenamefont {Su}, \citenamefont {Xi}, \citenamefont {Fu}, \citenamefont
		{Chen}, \citenamefont {Jin}, \citenamefont {Chen}, \citenamefont {Mo},
		\citenamefont {Qi}, \citenamefont {Shen}, \citenamefont {Zhang},
		\citenamefont {Jin}, \citenamefont {Li}, \citenamefont {Sun},\ and\
		\citenamefont {Su}}]{Xiang2023}%
	\BibitemOpen
	\bibfield  {author} {\bibinfo {author} {\bibfnamefont {J.}~\bibnamefont
			{Xiang}}, \bibinfo {author} {\bibfnamefont {C.}~\bibnamefont {Su}}, \bibinfo
		{author} {\bibfnamefont {N.}~\bibnamefont {Xi}}, \bibinfo {author}
		{\bibfnamefont {Z.}~\bibnamefont {Fu}}, \bibinfo {author} {\bibfnamefont
			{Z.}~\bibnamefont {Chen}}, \bibinfo {author} {\bibfnamefont {H.}~\bibnamefont
			{Jin}}, \bibinfo {author} {\bibfnamefont {Z.}~\bibnamefont {Chen}}, \bibinfo
		{author} {\bibfnamefont {Z.-J.}\ \bibnamefont {Mo}}, \bibinfo {author}
		{\bibfnamefont {Y.}~\bibnamefont {Qi}}, \bibinfo {author} {\bibfnamefont
			{J.}~\bibnamefont {Shen}}, \bibinfo {author} {\bibfnamefont {L.}~\bibnamefont
			{Zhang}}, \bibinfo {author} {\bibfnamefont {W.}~\bibnamefont {Jin}}, \bibinfo
		{author} {\bibfnamefont {W.}~\bibnamefont {Li}}, \bibinfo {author}
		{\bibfnamefont {P.}~\bibnamefont {Sun}},\ and\ \bibinfo {author}
		{\bibfnamefont {G.}~\bibnamefont {Su}},\ }\bibfield  {title} {\bibinfo
		{title} {{Dipolar Spin Liquid Ending with Quantum Critical Point in a
				Gd-based Triangular Magnet}},\ }\href@noop {} {\ }\Eprint
	{https://arxiv.org/abs/2301.03571} {arXiv:2301.03571 [cond-mat.str-el]}
	\BibitemShut {NoStop}%
	\bibitem [{\citenamefont {Kuznetsov}\ \emph {et~al.}(2022)\citenamefont
		{Kuznetsov}, \citenamefont {Kokh}, \citenamefont {Sagatov}, \citenamefont
		{Gavryushkin}, \citenamefont {Molokeev}, \citenamefont {Svetlichnyi},
		\citenamefont {Lapin}, \citenamefont {Kononova}, \citenamefont {Shevchenko},
		\citenamefont {Bolatov}, \citenamefont {Uralbekov}, \citenamefont
		{Goreiavcheva},\ and\ \citenamefont {Kokh}}]{Kuznetsov7497}%
	\BibitemOpen
	\bibfield  {author} {\bibinfo {author} {\bibfnamefont {A.~B.}\ \bibnamefont
			{Kuznetsov}}, \bibinfo {author} {\bibfnamefont {K.~A.}\ \bibnamefont {Kokh}},
		\bibinfo {author} {\bibfnamefont {N.}~\bibnamefont {Sagatov}}, \bibinfo
		{author} {\bibfnamefont {P.~N.}\ \bibnamefont {Gavryushkin}}, \bibinfo
		{author} {\bibfnamefont {M.~S.}\ \bibnamefont {Molokeev}}, \bibinfo {author}
		{\bibfnamefont {V.~A.}\ \bibnamefont {Svetlichnyi}}, \bibinfo {author}
		{\bibfnamefont {I.~N.}\ \bibnamefont {Lapin}}, \bibinfo {author}
		{\bibfnamefont {N.~G.}\ \bibnamefont {Kononova}}, \bibinfo {author}
		{\bibfnamefont {V.~S.}\ \bibnamefont {Shevchenko}}, \bibinfo {author}
		{\bibfnamefont {A.}~\bibnamefont {Bolatov}}, \bibinfo {author} {\bibfnamefont
			{B.}~\bibnamefont {Uralbekov}}, \bibinfo {author} {\bibfnamefont {A.~A.}\
			\bibnamefont {Goreiavcheva}},\ and\ \bibinfo {author} {\bibfnamefont {A.~E.}\
			\bibnamefont {Kokh}},\ }\bibfield  {title} {\bibinfo {title} {{Synthesis and
				Growth of Rare Earth Borates NaSr$R$(BO$_3$)$_2$ ($R$ = Ho--Lu, Y, Sc)}},\
	}\href {https://doi.org/10.1021/acs.inorgchem.2c00596} {\bibfield  {journal}
		{\bibinfo  {journal} {Inorg. Chem.}\ }\textbf {\bibinfo {volume} {61}},\
		\bibinfo {pages} {7497} (\bibinfo {year} {2022})}\BibitemShut {NoStop}%
	\bibitem [{\citenamefont {Dabi{\'{c}}}\ \emph {et~al.}(2021)\citenamefont
		{Dabi{\'{c}}}, \citenamefont {Kahlenberg}, \citenamefont {Kr{\"{u}}ger},
		\citenamefont {Rodi{\'{c}}}, \citenamefont {Kova{\v{c}}}, \citenamefont
		{Blanu{\v{s}}a}, \citenamefont {Jagli{\v{c}}i{\'{c}}}, \citenamefont
		{Karanovi{\'{c}}}, \citenamefont {Pet{\v{r}}{\'\i}{\v{c}}ek},\ and\
		\citenamefont {Kremenovi{\'{c}}}}]{Dabic584}%
	\BibitemOpen
	\bibfield  {author} {\bibinfo {author} {\bibfnamefont {P.}~\bibnamefont
			{Dabi{\'{c}}}}, \bibinfo {author} {\bibfnamefont {V.}~\bibnamefont
			{Kahlenberg}}, \bibinfo {author} {\bibfnamefont {B.}~\bibnamefont
			{Kr{\"{u}}ger}}, \bibinfo {author} {\bibfnamefont {M.}~\bibnamefont
			{Rodi{\'{c}}}}, \bibinfo {author} {\bibfnamefont {S.}~\bibnamefont
			{Kova{\v{c}}}}, \bibinfo {author} {\bibfnamefont {J.}~\bibnamefont
			{Blanu{\v{s}}a}}, \bibinfo {author} {\bibfnamefont {Z.}~\bibnamefont
			{Jagli{\v{c}}i{\'{c}}}}, \bibinfo {author} {\bibfnamefont {L.}~\bibnamefont
			{Karanovi{\'{c}}}}, \bibinfo {author} {\bibfnamefont {V.}~\bibnamefont
			{Pet{\v{r}}{\'\i}{\v{c}}ek}},\ and\ \bibinfo {author} {\bibfnamefont
			{A.}~\bibnamefont {Kremenovi{\'{c}}}},\ }\bibfield  {title} {\bibinfo {title}
		{{Low-temperature phase transition and magnetic properties of K${\sb
					3}$YbSi${\sb 2}$O${\sb 7}$}},\ }\href
	{https://doi.org/10.1107/S2052520621006077} {\bibfield  {journal} {\bibinfo
			{journal} {Acta Crystallogr. B}\ }\textbf {\bibinfo {volume} {77}},\ \bibinfo
		{pages} {584} (\bibinfo {year} {2021})}\BibitemShut {NoStop}%
	\bibitem [{\citenamefont {Rodríguez-Carvajal}(1993)}]{Carvajal55}%
	\BibitemOpen
	\bibfield  {author} {\bibinfo {author} {\bibfnamefont {J.}~\bibnamefont
			{Rodríguez-Carvajal}},\ }\bibfield  {title} {\bibinfo {title} {Recent
			advances in magnetic structure determination by neutron powder diffraction},\
	}\href {https://doi.org/https://doi.org/10.1016/0921-4526(93)90108-I}
	{\bibfield  {journal} {\bibinfo  {journal} {Physica B Condens. Matter}\
		}\textbf {\bibinfo {volume} {192}},\ \bibinfo {pages} {55} (\bibinfo {year}
		{1993})}\BibitemShut {NoStop}%
	\bibitem [{Sup()}]{Supplementary}%
	\BibitemOpen
	\href@noop {} {}\bibinfo {note} {See Supplemental Material at http: LINK for
		additional information}\BibitemShut {NoStop}%
	\bibitem [{\citenamefont {Scheie}(2021)}]{Scheie356}%
	\BibitemOpen
	\bibfield  {author} {\bibinfo {author} {\bibfnamefont {A.}~\bibnamefont
			{Scheie}},\ }\bibfield  {title} {\bibinfo {title} {{{\it PyCrystalField}:
				software for calculation, analysis and fitting of crystal electric field
				Hamiltonians}},\ }\href {https://doi.org/10.1107/S160057672001554X}
	{\bibfield  {journal} {\bibinfo  {journal} {J. Appl. Cryst.}\ }\textbf
		{\bibinfo {volume} {54}},\ \bibinfo {pages} {356} (\bibinfo {year}
		{2021})}\BibitemShut {NoStop}%
	\bibitem [{\citenamefont {Moon}\ \emph {et~al.}(1968)\citenamefont {Moon},
		\citenamefont {Koehler}, \citenamefont {Child},\ and\ \citenamefont
		{Raubenheimer}}]{Moon722}%
	\BibitemOpen
	\bibfield  {author} {\bibinfo {author} {\bibfnamefont {R.~M.}\ \bibnamefont
			{Moon}}, \bibinfo {author} {\bibfnamefont {W.~C.}\ \bibnamefont {Koehler}},
		\bibinfo {author} {\bibfnamefont {H.~R.}\ \bibnamefont {Child}},\ and\
		\bibinfo {author} {\bibfnamefont {L.~J.}\ \bibnamefont {Raubenheimer}},\
	}\bibfield  {title} {\bibinfo {title} {{Magnetic Structures of
				${\mathrm{Er}}_{2}$${\mathrm{O}}_{3}$ and
				${\mathrm{Yb}}_{2}$${\mathrm{O}}_{3}$}},\ }\href
	{https://doi.org/10.1103/PhysRev.176.722} {\bibfield  {journal} {\bibinfo
			{journal} {Phys. Rev.}\ }\textbf {\bibinfo {volume} {176}},\ \bibinfo {pages}
		{722} (\bibinfo {year} {1968})}\BibitemShut {NoStop}%
	\bibitem [{\citenamefont {Arjun}\ \emph
		{et~al.}(2023{\natexlab{a}})\citenamefont {Arjun}, \citenamefont {Ranjith},
		\citenamefont {Jesche}, \citenamefont {Hirschberger}, \citenamefont {Sarma},\
		and\ \citenamefont {Gegenwart}}]{Arjun224415}%
	\BibitemOpen
	\bibfield  {author} {\bibinfo {author} {\bibfnamefont {U.}~\bibnamefont
			{Arjun}}, \bibinfo {author} {\bibfnamefont {K.~M.}\ \bibnamefont {Ranjith}},
		\bibinfo {author} {\bibfnamefont {A.}~\bibnamefont {Jesche}}, \bibinfo
		{author} {\bibfnamefont {F.}~\bibnamefont {Hirschberger}}, \bibinfo {author}
		{\bibfnamefont {D.~D.}\ \bibnamefont {Sarma}},\ and\ \bibinfo {author}
		{\bibfnamefont {P.}~\bibnamefont {Gegenwart}},\ }\bibfield  {title} {\bibinfo
		{title} {{Adiabatic demagnetization refrigeration to millikelvin temperatures
				with the distorted square lattice magnet ${\mathrm{NaYbGeO}}_{4}$}},\ }\href
	{https://doi.org/10.1103/PhysRevB.108.224415} {\bibfield  {journal} {\bibinfo
			{journal} {Phys. Rev. B}\ }\textbf {\bibinfo {volume} {108}},\ \bibinfo
		{pages} {224415} (\bibinfo {year} {2023}{\natexlab{a}})}\BibitemShut
	{NoStop}%
	\bibitem [{\citenamefont {Guchhait}\ \emph {et~al.}(2024)\citenamefont
		{Guchhait}, \citenamefont {Painganoor}, \citenamefont {Islam}, \citenamefont
		{Sichelschmidt}, \citenamefont {Le}, \citenamefont {Christensen},\ and\
		\citenamefont {Nath}}]{Guchhait144434}%
	\BibitemOpen
	\bibfield  {author} {\bibinfo {author} {\bibfnamefont {S.}~\bibnamefont
			{Guchhait}}, \bibinfo {author} {\bibfnamefont {A.}~\bibnamefont
			{Painganoor}}, \bibinfo {author} {\bibfnamefont {S.~S.}\ \bibnamefont
			{Islam}}, \bibinfo {author} {\bibfnamefont {J.}~\bibnamefont
			{Sichelschmidt}}, \bibinfo {author} {\bibfnamefont {M.~D.}\ \bibnamefont
			{Le}}, \bibinfo {author} {\bibfnamefont {N.~B.}\ \bibnamefont
			{Christensen}},\ and\ \bibinfo {author} {\bibfnamefont {R.}~\bibnamefont
			{Nath}},\ }\bibfield  {title} {\bibinfo {title} {{Magnetic and crystal
				electric field studies of the rare earth based square lattice antiferromagnet
				${\mathrm{NdKNaNbO}}_{5}$}},\ }\href
	{https://doi.org/10.1103/PhysRevB.110.144434} {\bibfield  {journal} {\bibinfo
			{journal} {Phys. Rev. B}\ }\textbf {\bibinfo {volume} {110}},\ \bibinfo
		{pages} {144434} (\bibinfo {year} {2024})}\BibitemShut {NoStop}%
	\bibitem [{\citenamefont {Somesh}\ \emph {et~al.}(2023)\citenamefont {Somesh},
		\citenamefont {Islam}, \citenamefont {Mohanty}, \citenamefont {Simutis},
		\citenamefont {Guguchia}, \citenamefont {Wang}, \citenamefont
		{Sichelschmidt}, \citenamefont {Baenitz},\ and\ \citenamefont
		{Nath}}]{Somesh064421}%
	\BibitemOpen
	\bibfield  {author} {\bibinfo {author} {\bibfnamefont {K.}~\bibnamefont
			{Somesh}}, \bibinfo {author} {\bibfnamefont {S.~S.}\ \bibnamefont {Islam}},
		\bibinfo {author} {\bibfnamefont {S.}~\bibnamefont {Mohanty}}, \bibinfo
		{author} {\bibfnamefont {G.}~\bibnamefont {Simutis}}, \bibinfo {author}
		{\bibfnamefont {Z.}~\bibnamefont {Guguchia}}, \bibinfo {author}
		{\bibfnamefont {C.}~\bibnamefont {Wang}}, \bibinfo {author} {\bibfnamefont
			{J.}~\bibnamefont {Sichelschmidt}}, \bibinfo {author} {\bibfnamefont
			{M.}~\bibnamefont {Baenitz}},\ and\ \bibinfo {author} {\bibfnamefont
			{R.}~\bibnamefont {Nath}},\ }\bibfield  {title} {\bibinfo {title} {{Absence
				of magnetic order and emergence of unconventional fluctuations in the $J_{\rm
					eff}=\frac{1}{2}$ triangular-lattice antiferromagnet YbBO$_{3}$}},\ }\href
	{https://doi.org/10.1103/PhysRevB.107.064421} {\bibfield  {journal} {\bibinfo
			{journal} {Phys. Rev. B}\ }\textbf {\bibinfo {volume} {107}},\ \bibinfo
		{pages} {064421} (\bibinfo {year} {2023})}\BibitemShut {NoStop}%
	\bibitem [{\citenamefont {Ranjith}\ \emph {et~al.}(2017)\citenamefont
		{Ranjith}, \citenamefont {Brinda}, \citenamefont {Arjun}, \citenamefont
		{Hegde},\ and\ \citenamefont {Nath}}]{Ranjith115804}%
	\BibitemOpen
	\bibfield  {author} {\bibinfo {author} {\bibfnamefont {K.~M.}\ \bibnamefont
			{Ranjith}}, \bibinfo {author} {\bibfnamefont {K.}~\bibnamefont {Brinda}},
		\bibinfo {author} {\bibfnamefont {U.}~\bibnamefont {Arjun}}, \bibinfo
		{author} {\bibfnamefont {N.~G.}\ \bibnamefont {Hegde}},\ and\ \bibinfo
		{author} {\bibfnamefont {R.}~\bibnamefont {Nath}},\ }\bibfield  {title}
	{\bibinfo {title} {{Double phase transition in the triangular antiferromagnet
				Ba$_3$CoTa$_2$O$_9$}},\ }\href {https://doi.org/10.1088/1361-648X/aa57be}
	{\bibfield  {journal} {\bibinfo  {journal} {J. Phys.: Condens. Matter}\
		}\textbf {\bibinfo {volume} {29}},\ \bibinfo {pages} {115804} (\bibinfo
		{year} {2017})}\BibitemShut {NoStop}%
	\bibitem [{\citenamefont {Ranjith}\ \emph
		{et~al.}(2019{\natexlab{a}})\citenamefont {Ranjith}, \citenamefont
		{Dmytriieva}, \citenamefont {Khim}, \citenamefont {Sichelschmidt},
		\citenamefont {Luther}, \citenamefont {Ehlers}, \citenamefont {Yasuoka},
		\citenamefont {Wosnitza}, \citenamefont {Tsirlin}, \citenamefont {K\"uhne},\
		and\ \citenamefont {Baenitz}}]{Ranjith180401}%
	\BibitemOpen
	\bibfield  {author} {\bibinfo {author} {\bibfnamefont {K.~M.}\ \bibnamefont
			{Ranjith}}, \bibinfo {author} {\bibfnamefont {D.}~\bibnamefont {Dmytriieva}},
		\bibinfo {author} {\bibfnamefont {S.}~\bibnamefont {Khim}}, \bibinfo {author}
		{\bibfnamefont {J.}~\bibnamefont {Sichelschmidt}}, \bibinfo {author}
		{\bibfnamefont {S.}~\bibnamefont {Luther}}, \bibinfo {author} {\bibfnamefont
			{D.}~\bibnamefont {Ehlers}}, \bibinfo {author} {\bibfnamefont
			{H.}~\bibnamefont {Yasuoka}}, \bibinfo {author} {\bibfnamefont
			{J.}~\bibnamefont {Wosnitza}}, \bibinfo {author} {\bibfnamefont {A.~A.}\
			\bibnamefont {Tsirlin}}, \bibinfo {author} {\bibfnamefont {H.}~\bibnamefont
			{K\"uhne}},\ and\ \bibinfo {author} {\bibfnamefont {M.}~\bibnamefont
			{Baenitz}},\ }\bibfield  {title} {\bibinfo {title} {{Field-induced
				instability of the quantum spin liquid ground state in the
				${J}_{\mathrm{eff}}=\frac{1}{2}$ triangular-lattice compound
				${\mathrm{NaYbO}}_{2}$}},\ }\href
	{https://doi.org/10.1103/PhysRevB.99.180401} {\bibfield  {journal} {\bibinfo
			{journal} {Phys. Rev. B}\ }\textbf {\bibinfo {volume} {99}},\ \bibinfo
		{pages} {180401} (\bibinfo {year} {2019}{\natexlab{a}})}\BibitemShut
	{NoStop}%
	\bibitem [{\citenamefont {Ranjith}\ \emph
		{et~al.}(2019{\natexlab{b}})\citenamefont {Ranjith}, \citenamefont {Luther},
		\citenamefont {Reimann}, \citenamefont {Schmidt}, \citenamefont {Schlender},
		\citenamefont {Sichelschmidt}, \citenamefont {Yasuoka}, \citenamefont
		{Strydom}, \citenamefont {Skourski}, \citenamefont {Wosnitza}, \citenamefont
		{K\"uhne}, \citenamefont {Doert},\ and\ \citenamefont
		{Baenitz}}]{Ranjith224417}%
	\BibitemOpen
	\bibfield  {author} {\bibinfo {author} {\bibfnamefont {K.~M.}\ \bibnamefont
			{Ranjith}}, \bibinfo {author} {\bibfnamefont {S.}~\bibnamefont {Luther}},
		\bibinfo {author} {\bibfnamefont {T.}~\bibnamefont {Reimann}}, \bibinfo
		{author} {\bibfnamefont {B.}~\bibnamefont {Schmidt}}, \bibinfo {author}
		{\bibfnamefont {P.}~\bibnamefont {Schlender}}, \bibinfo {author}
		{\bibfnamefont {J.}~\bibnamefont {Sichelschmidt}}, \bibinfo {author}
		{\bibfnamefont {H.}~\bibnamefont {Yasuoka}}, \bibinfo {author} {\bibfnamefont
			{A.~M.}\ \bibnamefont {Strydom}}, \bibinfo {author} {\bibfnamefont
			{Y.}~\bibnamefont {Skourski}}, \bibinfo {author} {\bibfnamefont
			{J.}~\bibnamefont {Wosnitza}}, \bibinfo {author} {\bibfnamefont
			{H.}~\bibnamefont {K\"uhne}}, \bibinfo {author} {\bibfnamefont
			{T.}~\bibnamefont {Doert}},\ and\ \bibinfo {author} {\bibfnamefont
			{M.}~\bibnamefont {Baenitz}},\ }\bibfield  {title} {\bibinfo {title}
		{{Anisotropic field-induced ordering in the triangular-lattice quantum spin
				liquid ${\mathrm{NaYbSe}}_{2}$}},\ }\href
	{https://doi.org/10.1103/PhysRevB.100.224417} {\bibfield  {journal} {\bibinfo
			{journal} {Phys. Rev. B}\ }\textbf {\bibinfo {volume} {100}},\ \bibinfo
		{pages} {224417} (\bibinfo {year} {2019}{\natexlab{b}})}\BibitemShut
	{NoStop}%
	\bibitem [{\citenamefont {Mugiraneza}\ and\ \citenamefont
		{Hallas}(2022)}]{Mugiraneza95}%
	\BibitemOpen
	\bibfield  {author} {\bibinfo {author} {\bibfnamefont {S.}~\bibnamefont
			{Mugiraneza}}\ and\ \bibinfo {author} {\bibfnamefont {A.~M.}\ \bibnamefont
			{Hallas}},\ }\bibfield  {title} {\bibinfo {title} {{Tutorial: a beginner's
				guide to interpreting magnetic susceptibility data with the Curie-Weiss
				law}},\ }\href {https://doi.org/10.1038/s42005-022-00853-y} {\bibfield
		{journal} {\bibinfo  {journal} {Commun. Phys.}\ }\textbf {\bibinfo {volume}
			{5}},\ \bibinfo {pages} {95} (\bibinfo {year} {2022})}\BibitemShut {NoStop}%
	\bibitem [{\citenamefont {Arjun}\ \emph
		{et~al.}(2023{\natexlab{b}})\citenamefont {Arjun}, \citenamefont {Ranjith},
		\citenamefont {Jesche}, \citenamefont {Hirschberger}, \citenamefont {Sarma},\
		and\ \citenamefont {Gegenwart}}]{Arjun014013}%
	\BibitemOpen
	\bibfield  {author} {\bibinfo {author} {\bibfnamefont {U.}~\bibnamefont
			{Arjun}}, \bibinfo {author} {\bibfnamefont {K.}~\bibnamefont {Ranjith}},
		\bibinfo {author} {\bibfnamefont {A.}~\bibnamefont {Jesche}}, \bibinfo
		{author} {\bibfnamefont {F.}~\bibnamefont {Hirschberger}}, \bibinfo {author}
		{\bibfnamefont {D.}~\bibnamefont {Sarma}},\ and\ \bibinfo {author}
		{\bibfnamefont {P.}~\bibnamefont {Gegenwart}},\ }\bibfield  {title} {\bibinfo
		{title} {{Efficient Adiabatic Demagnetization Refrigeration to below 50 mK
				with Ultrahigh-Vacuum-Compatible Ytterbium Diphosphates
				$A{\mathrm{YbP}}_{2}{\mathrm{O}}_{7}$ ($A$=Na, K)}},\ }\href
	{https://doi.org/10.1103/PhysRevApplied.20.014013} {\bibfield  {journal}
		{\bibinfo  {journal} {Phys. Rev. Appl.}\ }\textbf {\bibinfo {volume} {20}},\
		\bibinfo {pages} {014013} (\bibinfo {year} {2023}{\natexlab{b}})}\BibitemShut
	{NoStop}%
	\bibitem [{\citenamefont {Pula}\ \emph {et~al.}(2024)\citenamefont {Pula},
		\citenamefont {Sharma}, \citenamefont {Gautreau}, \citenamefont {K.~P.},
		\citenamefont {Kanigel}, \citenamefont {Frontzek}, \citenamefont {Dolling},
		\citenamefont {Clark}, \citenamefont {Dunsiger}, \citenamefont {Ghara},\ and\
		\citenamefont {Luke}}]{Pula014412}%
	\BibitemOpen
	\bibfield  {author} {\bibinfo {author} {\bibfnamefont {M.}~\bibnamefont
			{Pula}}, \bibinfo {author} {\bibfnamefont {S.}~\bibnamefont {Sharma}},
		\bibinfo {author} {\bibfnamefont {J.}~\bibnamefont {Gautreau}}, \bibinfo
		{author} {\bibfnamefont {S.}~\bibnamefont {K.~P.}}, \bibinfo {author}
		{\bibfnamefont {A.}~\bibnamefont {Kanigel}}, \bibinfo {author} {\bibfnamefont
			{M.~D.}\ \bibnamefont {Frontzek}}, \bibinfo {author} {\bibfnamefont {T.~N.}\
			\bibnamefont {Dolling}}, \bibinfo {author} {\bibfnamefont {L.}~\bibnamefont
			{Clark}}, \bibinfo {author} {\bibfnamefont {S.}~\bibnamefont {Dunsiger}},
		\bibinfo {author} {\bibfnamefont {A.}~\bibnamefont {Ghara}},\ and\ \bibinfo
		{author} {\bibfnamefont {G.~M.}\ \bibnamefont {Luke}},\ }\bibfield  {title}
	{\bibinfo {title} {{Candidate for a quantum spin liquid ground state in the
				Shastry-Sutherland lattice material
				${\mathrm{Yb}}_{2}{\mathrm{Be}}_{2}{\mathrm{GeO}}_{7}$}},\ }\href
	{https://doi.org/10.1103/PhysRevB.110.014412} {\bibfield  {journal} {\bibinfo
			{journal} {Phys. Rev. B}\ }\textbf {\bibinfo {volume} {110}},\ \bibinfo
		{pages} {014412} (\bibinfo {year} {2024})}\BibitemShut {NoStop}%
	\bibitem [{\citenamefont {Li}\ \emph {et~al.}(2015)\citenamefont {Li},
		\citenamefont {Chen}, \citenamefont {Tong}, \citenamefont {Pi}, \citenamefont
		{Liu}, \citenamefont {Yang}, \citenamefont {Wang},\ and\ \citenamefont
		{Zhang}}]{Li167203}%
	\BibitemOpen
	\bibfield  {author} {\bibinfo {author} {\bibfnamefont {Y.}~\bibnamefont
			{Li}}, \bibinfo {author} {\bibfnamefont {G.}~\bibnamefont {Chen}}, \bibinfo
		{author} {\bibfnamefont {W.}~\bibnamefont {Tong}}, \bibinfo {author}
		{\bibfnamefont {L.}~\bibnamefont {Pi}}, \bibinfo {author} {\bibfnamefont
			{J.}~\bibnamefont {Liu}}, \bibinfo {author} {\bibfnamefont {Z.}~\bibnamefont
			{Yang}}, \bibinfo {author} {\bibfnamefont {X.}~\bibnamefont {Wang}},\ and\
		\bibinfo {author} {\bibfnamefont {Q.}~\bibnamefont {Zhang}},\ }\bibfield
	{title} {\bibinfo {title} {{Rare-Earth Triangular Lattice Spin Liquid: A
				Single-Crystal Study of ${\mathrm{YbMgGaO}}_{4}$}},\ }\href
	{https://doi.org/10.1103/PhysRevLett.115.167203} {\bibfield  {journal}
		{\bibinfo  {journal} {Phys. Rev. Lett.}\ }\textbf {\bibinfo {volume} {115}},\
		\bibinfo {pages} {167203} (\bibinfo {year} {2015})}\BibitemShut {NoStop}%
	\bibitem [{\citenamefont {Thamban}\ \emph {et~al.}(2017)\citenamefont
		{Thamban}, \citenamefont {Arjun}, \citenamefont {Padmanabhan},\ and\
		\citenamefont {Nath}}]{Thamban255801}%
	\BibitemOpen
	\bibfield  {author} {\bibinfo {author} {\bibfnamefont {S.}~\bibnamefont
			{Thamban}}, \bibinfo {author} {\bibfnamefont {U.}~\bibnamefont {Arjun}},
		\bibinfo {author} {\bibfnamefont {M.}~\bibnamefont {Padmanabhan}},\ and\
		\bibinfo {author} {\bibfnamefont {R.}~\bibnamefont {Nath}},\ }\bibfield
	{title} {\bibinfo {title} {{Structural and magnetic properties of spin-1/2
				dimer compound Cu2(IPA)2(DMF)(H2O) with a large spin gap}},\ }\href
	{https://doi.org/10.1088/1361-648X/aa6ecb} {\bibfield  {journal} {\bibinfo
			{journal} {J. Phys.: Cond. Mat.}\ }\textbf {\bibinfo {volume} {29}},\
		\bibinfo {pages} {255801} (\bibinfo {year} {2017})}\BibitemShut {NoStop}%
	\bibitem [{\citenamefont {Biswal}\ \emph {et~al.}(2023)\citenamefont {Biswal},
		\citenamefont {Guchhait}, \citenamefont {Ghosh}, \citenamefont {Sarangi},
		\citenamefont {Samal}, \citenamefont {Swain}, \citenamefont {Kumar},\ and\
		\citenamefont {Nath}}]{Biswal134420}%
	\BibitemOpen
	\bibfield  {author} {\bibinfo {author} {\bibfnamefont {P.}~\bibnamefont
			{Biswal}}, \bibinfo {author} {\bibfnamefont {S.}~\bibnamefont {Guchhait}},
		\bibinfo {author} {\bibfnamefont {S.}~\bibnamefont {Ghosh}}, \bibinfo
		{author} {\bibfnamefont {S.~N.}\ \bibnamefont {Sarangi}}, \bibinfo {author}
		{\bibfnamefont {D.}~\bibnamefont {Samal}}, \bibinfo {author} {\bibfnamefont
			{D.}~\bibnamefont {Swain}}, \bibinfo {author} {\bibfnamefont
			{M.}~\bibnamefont {Kumar}},\ and\ \bibinfo {author} {\bibfnamefont
			{R.}~\bibnamefont {Nath}},\ }\bibfield  {title} {\bibinfo {title} {{Crystal
				structure and magnetic properties of the spin-$\frac{1}{2}$ frustrated
				two-leg ladder compounds
				$({\mathrm{C}}_{4}{\mathrm{H}}_{14}{\mathrm{N}}_{2}){\mathrm{Cu}}_{2}{X}_{6}$
				$(X = \mathrm{Cl} \text{and} \mathrm{Br})$}},\ }\href
	{https://doi.org/10.1103/PhysRevB.108.134420} {\bibfield  {journal} {\bibinfo
			{journal} {Phys. Rev. B}\ }\textbf {\bibinfo {volume} {108}},\ \bibinfo
		{pages} {134420} (\bibinfo {year} {2023})}\BibitemShut {NoStop}%
	\bibitem [{\citenamefont {Kittel}(1986)}]{Kittel1986}%
	\BibitemOpen
	\bibfield  {author} {\bibinfo {author} {\bibfnamefont {C.}~\bibnamefont
			{Kittel}},\ }\href@noop {} {\emph {\bibinfo {title} {{Introduction to Solid
					State Physics}}}},\ \bibinfo {edition} {8th}\ ed.\ (\bibinfo  {publisher}
	{John Wiley \& Sons, Inc.},\ \bibinfo {address} {New York},\ \bibinfo {year}
	{1986})\BibitemShut {NoStop}%
	\bibitem [{\citenamefont {Nath}\ \emph {et~al.}(2015)\citenamefont {Nath},
		\citenamefont {Padmanabhan}, \citenamefont {Baby}, \citenamefont
		{Thirumurugan}, \citenamefont {Ehlers}, \citenamefont {Hemmida},
		\citenamefont {Krug~von Nidda},\ and\ \citenamefont {Tsirlin}}]{Nath054409}%
	\BibitemOpen
	\bibfield  {author} {\bibinfo {author} {\bibfnamefont {R.}~\bibnamefont
			{Nath}}, \bibinfo {author} {\bibfnamefont {M.}~\bibnamefont {Padmanabhan}},
		\bibinfo {author} {\bibfnamefont {S.}~\bibnamefont {Baby}}, \bibinfo {author}
		{\bibfnamefont {A.}~\bibnamefont {Thirumurugan}}, \bibinfo {author}
		{\bibfnamefont {D.}~\bibnamefont {Ehlers}}, \bibinfo {author} {\bibfnamefont
			{M.}~\bibnamefont {Hemmida}}, \bibinfo {author} {\bibfnamefont {H.-A.}\
			\bibnamefont {Krug~von Nidda}},\ and\ \bibinfo {author} {\bibfnamefont
			{A.~A.}\ \bibnamefont {Tsirlin}},\ }\bibfield  {title} {\bibinfo {title}
		{{Quasi-two-dimensional $S=\frac{1}{2}$ magnetism of
				$\mathrm{Cu[}{\mathrm{C}}_{6}{\mathrm{H}}_{2}(\text{COO}{)}_{4}\mathrm{][}{\mathrm{C}}_{2}{\mathrm{H}}_{5}{\mathrm{NH}}_{3}\mathrm{]}{}_{2}$}},\
	}\href {https://doi.org/10.1103/PhysRevB.91.054409} {\bibfield  {journal}
		{\bibinfo  {journal} {Phys. Rev. B}\ }\textbf {\bibinfo {volume} {91}},\
		\bibinfo {pages} {054409} (\bibinfo {year} {2015})}\BibitemShut {NoStop}%
	\bibitem [{\citenamefont {Guo}\ \emph {et~al.}(2019)\citenamefont {Guo},
		\citenamefont {Ghasemi}, \citenamefont {Broholm},\ and\ \citenamefont
		{Cava}}]{Guo094404}%
	\BibitemOpen
	\bibfield  {author} {\bibinfo {author} {\bibfnamefont {S.}~\bibnamefont
			{Guo}}, \bibinfo {author} {\bibfnamefont {A.}~\bibnamefont {Ghasemi}},
		\bibinfo {author} {\bibfnamefont {C.~L.}\ \bibnamefont {Broholm}},\ and\
		\bibinfo {author} {\bibfnamefont {R.~J.}\ \bibnamefont {Cava}},\ }\bibfield
	{title} {\bibinfo {title} {Magnetism on ideal triangular lattices in
			$\mathrm{NaBaYb}{(\mathrm{B}{\mathrm{O}}_{3})}_{2}$},\ }\href
	{https://doi.org/10.1103/PhysRevMaterials.3.094404} {\bibfield  {journal}
		{\bibinfo  {journal} {Phys. Rev. Mater.}\ }\textbf {\bibinfo {volume} {3}},\
		\bibinfo {pages} {094404} (\bibinfo {year} {2019})}\BibitemShut {NoStop}%
	\bibitem [{\citenamefont {Newman}\ and\ \citenamefont
		{Ng}(2000)}]{Newman_Ng_2000}%
	\BibitemOpen
	\bibfield  {author} {\bibinfo {author} {\bibfnamefont {D.~J.}\ \bibnamefont
			{Newman}}\ and\ \bibinfo {author} {\bibfnamefont {B.}~\bibnamefont {Ng}},\
	}\href {https://doi.org/https://doi.org/10.1017/CBO9780511524295} {\emph
		{\bibinfo {title} {Crystal Field Handbook}}}\ (\bibinfo  {publisher}
	{Cambridge University Press},\ \bibinfo {year} {2000})\BibitemShut {NoStop}%
	\bibitem [{\citenamefont {Stevens}(1952)}]{Stevens209}%
	\BibitemOpen
	\bibfield  {author} {\bibinfo {author} {\bibfnamefont {K.~W.~H.}\
			\bibnamefont {Stevens}},\ }\bibfield  {title} {\bibinfo {title} {Matrix
			elements and operator equivalents connected with the magnetic properties of
			rare earth ions},\ }\href {https://doi.org/10.1088/0370-1298/65/3/308}
	{\bibfield  {journal} {\bibinfo  {journal} {Proc. Phys. Soc. Section A}\
		}\textbf {\bibinfo {volume} {65}},\ \bibinfo {pages} {209} (\bibinfo {year}
		{1952})}\BibitemShut {NoStop}%
	\bibitem [{\citenamefont {Hutchings}(1964)}]{Huthchings227}%
	\BibitemOpen
	\bibfield  {author} {\bibinfo {author} {\bibfnamefont {M.}~\bibnamefont
			{Hutchings}},\ }\href
	{https://doi.org/https://doi.org/10.1016/S0081-1947(08)60517-2} {\emph
		{\bibinfo {title} {Point-Charge Calculations of Energy Levels of Magnetic
				Ions in Crystalline Electric Fields}}},\ \bibinfo {series} {Solid State
		Physics}, Vol.~\bibinfo {volume} {16}\ (\bibinfo  {publisher} {Academic
		Press},\ \bibinfo {year} {1964})\ p.\ \bibinfo {pages} {227}\BibitemShut
	{NoStop}%
	\bibitem [{\citenamefont {Mohanty}\ \emph {et~al.}(2023)\citenamefont
		{Mohanty}, \citenamefont {Islam}, \citenamefont {Winterhalter-Stocker},
		\citenamefont {Jesche}, \citenamefont {Simutis}, \citenamefont {Wang},
		\citenamefont {Guguchia}, \citenamefont {Sichelschmidt}, \citenamefont
		{Baenitz}, \citenamefont {Tsirlin}, \citenamefont {Gegenwart},\ and\
		\citenamefont {Nath}}]{Mohanty134408}%
	\BibitemOpen
	\bibfield  {author} {\bibinfo {author} {\bibfnamefont {S.}~\bibnamefont
			{Mohanty}}, \bibinfo {author} {\bibfnamefont {S.~S.}\ \bibnamefont {Islam}},
		\bibinfo {author} {\bibfnamefont {N.}~\bibnamefont {Winterhalter-Stocker}},
		\bibinfo {author} {\bibfnamefont {A.}~\bibnamefont {Jesche}}, \bibinfo
		{author} {\bibfnamefont {G.}~\bibnamefont {Simutis}}, \bibinfo {author}
		{\bibfnamefont {C.}~\bibnamefont {Wang}}, \bibinfo {author} {\bibfnamefont
			{Z.}~\bibnamefont {Guguchia}}, \bibinfo {author} {\bibfnamefont
			{J.}~\bibnamefont {Sichelschmidt}}, \bibinfo {author} {\bibfnamefont
			{M.}~\bibnamefont {Baenitz}}, \bibinfo {author} {\bibfnamefont {A.~A.}\
			\bibnamefont {Tsirlin}}, \bibinfo {author} {\bibfnamefont {P.}~\bibnamefont
			{Gegenwart}},\ and\ \bibinfo {author} {\bibfnamefont {R.}~\bibnamefont
			{Nath}},\ }\bibfield  {title} {\bibinfo {title} {{Disordered ground state in
				the spin-orbit coupled ${J}_{\mathrm{eff}}$ = $\frac{1}{2}$ distorted
				honeycomb magnet BiYbGeO$_{5}$}},\ }\href
	{https://doi.org/10.1103/PhysRevB.108.134408} {\bibfield  {journal} {\bibinfo
			{journal} {Phys. Rev. B}\ }\textbf {\bibinfo {volume} {108}},\ \bibinfo
		{pages} {134408} (\bibinfo {year} {2023})}\BibitemShut {NoStop}%
	\bibitem [{\citenamefont {Xiang}\ \emph {et~al.}(2011)\citenamefont {Xiang},
		\citenamefont {Kan}, \citenamefont {Wei}, \citenamefont {Whangbo},\ and\
		\citenamefont {Gong}}]{Xiang224429}%
	\BibitemOpen
	\bibfield  {author} {\bibinfo {author} {\bibfnamefont {H.~J.}\ \bibnamefont
			{Xiang}}, \bibinfo {author} {\bibfnamefont {E.~J.}\ \bibnamefont {Kan}},
		\bibinfo {author} {\bibfnamefont {S.-H.}\ \bibnamefont {Wei}}, \bibinfo
		{author} {\bibfnamefont {M.-H.}\ \bibnamefont {Whangbo}},\ and\ \bibinfo
		{author} {\bibfnamefont {X.~G.}\ \bibnamefont {Gong}},\ }\bibfield  {title}
	{\bibinfo {title} {Predicting the spin-lattice order of frustrated systems
			from first principles},\ }\href {https://doi.org/10.1103/PhysRevB.84.224429}
	{\bibfield  {journal} {\bibinfo  {journal} {Phys. Rev. B}\ }\textbf {\bibinfo
			{volume} {84}},\ \bibinfo {pages} {224429} (\bibinfo {year}
		{2011})}\BibitemShut {NoStop}%
	\bibitem [{\citenamefont {Sebastian}\ \emph {et~al.}(2022)\citenamefont
		{Sebastian}, \citenamefont {Islam}, \citenamefont {Jain}, \citenamefont
		{Yusuf}, \citenamefont {Uhlarz},\ and\ \citenamefont
		{Nath}}]{Sebastian104425}%
	\BibitemOpen
	\bibfield  {author} {\bibinfo {author} {\bibfnamefont {S.~J.}\ \bibnamefont
			{Sebastian}}, \bibinfo {author} {\bibfnamefont {S.~S.}\ \bibnamefont
			{Islam}}, \bibinfo {author} {\bibfnamefont {A.}~\bibnamefont {Jain}},
		\bibinfo {author} {\bibfnamefont {S.~M.}\ \bibnamefont {Yusuf}}, \bibinfo
		{author} {\bibfnamefont {M.}~\bibnamefont {Uhlarz}},\ and\ \bibinfo {author}
		{\bibfnamefont {R.}~\bibnamefont {Nath}},\ }\bibfield  {title} {\bibinfo
		{title} {{Collinear order in the spin-$\frac{5}{2}$ triangular-lattice
				antiferromagnet ${\mathrm{Na}}_{3}\mathrm{Fe}{({\mathrm{PO}}_{4})}_{2}$}},\
	}\href {https://doi.org/10.1103/PhysRevB.105.104425} {\bibfield  {journal}
		{\bibinfo  {journal} {Phys. Rev. B}\ }\textbf {\bibinfo {volume} {105}},\
		\bibinfo {pages} {104425} (\bibinfo {year} {2022})}\BibitemShut {NoStop}%
\end{thebibliography}
								  
\end{document}